# Centromere reference models for human chromosomes X and Y satellite arrays

**SHORT TITLE:** Linear sequence models of human centromeric DNA

**Karen H. Miga[1,2], Yulia Newton[2], Miten Jain[2], Nicolas Altemose[1], Huntington F. Willard[1] and W. James Kent[2]***

[1] Duke Institute for Genome Sciences & Policy, Duke University, Durham, North Carolina, United States of America.

[2] Center for Biomolecular Science & Engineering, University of California Santa Cruz, Santa Cruz, California, United States of America

*Address for Correspondence:
W. James Kent
(831) 459-1401
Center for Biomolecular Science & Engineering
University of California
MS: CBSE-ITI
1156 High Street
Santa Cruz, CA 95064
kent@soe.ucsc.edu

**ABSTRACT**

The human genome sequence remains incomplete, with multi-megabase-sized gaps representing the endogenous centromeres and other heterochromatic regions. Available sequence-based studies within these sites in the genome have demonstrated a role in centromere function and chromosome pairing, necessary to ensure proper chromosome segregation during cell division.  A common genomic feature of these regions is the enrichment of long arrays of near-identical tandem repeats, known as satellite DNAs, that offer a limited number of variant sites to differentiate individual repeat copies across millions of bases. This substantial sequence homogeneity challenges available assembly strategies and, as a result, centromeric regions are omitted from ongoing genomic studies. To address this problem, we utilize monomer sequence and ordering information obtained from whole-genome shotgun reads to model two haploid human satellite arrays on chromosomes X and Y, resulting in an initial characterization of 3.83 Mb of centromeric DNA within an individual genome. To further expand the utility of each centromeric reference sequence models, we evaluate sites within the arrays for short-read mappability and chromosome specificity.  Because satellite DNAs evolve in a concerted manner, we use these centromeric assemblies to assess the extent of sequence variation among 372 individuals from distinct human populations. We thus identify two ancient satellite array variants in both X and Y centromeres, as determined by array length and sequence composition. This study provides an initial sequence characterization of a regional centromere and establishes a foundation to extend genomic characterization to these sites as well as to other repeat-rich regions within complex genomes.

**LINKS**

BioProject ID PRJNA193213; GenBank accession numbers: GK000058 (gi 529053718) and GK000059 (gi 529053714). Prior to the release of the GRCh38 reference, DXZ1 and DYZ3 sequence submissions have been made available from the following website: http://hgwdev.soe.ucsc.edu/~kehayden/GenomeResearchSeq/GK000058.fa (cenX) and GK000059.fa (cenY).

All documentation and source code for the linearSat is freely available at github.com/JimKent/linearSat.

**INTRODUCTION**

Extensive tracts of near-identical tandem repeats, known as satellite DNA arrays, are associated with constitutive heterochromatin and commonly provide the sequence definition for regional centromeres, or sites responsible for chromosome segregation (Yunis and Yasmineh 1971; Willard 1990; Schueler et al. 2001). Proper regulation of these sites is critical for cellular viability, as disruption in epigenetic maintenance often leads to genome instability and aneuploidy (Dernburg et al. 1996; Peng and Karpen 2008; Ting et al. 2011; Zhu et al. 2011). Despite their biological importance, the full sequence definition of satellite DNA-rich regions remains incomplete and largely uncharacterized even within extensively sequenced and studied genomes, resulting in large, multi-megabase gaps within each chromosome assembly (Henikoff 2002; Eichler et al. 2004; Rudd and Willard 2004). This presents a fundamental challenge to ongoing genomic studies aimed at understanding the role of these specialized domains in cellular function and emphasizes the need for a more complete representation of sequences that comprise highly homogenized arrays.

Efforts to predict the linear sequence arrangement within satellite arrays are impeded by insufficient and sparsely arranged sites capable of distinguishing one copy of the repeat from another, resulting in an increase in read coverage and assembly collapse (Durfy and Willard 1989; Schueler et al. 2001; Schindelhauer and Schwarz 2002; Treangen and Salzberg 2011). The extreme homogeneity further exacerbates the challenges faced by BAC-based sequence assembly, resulting in only marginal satellite DNA directly adjacent to heterochromatic and centromeric gaps in most reference assemblies (Eichler et al. 2004; Rudd and Willard 2004). Although these variant sites present a methodological challenge, their representation is necessary to study mechanisms of array evolution (Willard and Waye 1987; Santos et al. 1995; Warburton and Willard 1995), improve long-range physical maps (Wevrick and Willard 1989; Luce et al. 2006), and map sites of epigenetic enrichment that are important for centromere function (Maloney et al. 2012; Hayden et al. 2013). Thus, much effort has been devoted to designing approaches that will decipher satellite arrays.

Indeed, extensive experimental studies of human centromeric satellite arrays have provided a foundation on which to extend characterization of array sequence composition, organization, and evolution (Willard et al. 1989). Endogenous centromeres in the human genome are enriched with a single AT-rich satellite family, known as alpha satellite (Manuelidis 1978). Inherent sequence diversity between copies of the fundamental ~171-bp repeat unit enables alpha satellite monomer organization to be studied in a chromosome-specific manner (Willard 1985). High-resolution genomic characterization of alpha satellite organization reveals two general subtypes: those that appear to be highly divergent with few apparent local homology patterns, known as monomeric, and those monomers that are organized into tandemly-repeated, multi-monomer units, known as higher-order repeat (HOR) units (Willard and Waye 1987; Alexandrov et al. 1993). In contrast to monomeric alpha satellite, where near-identical tandem repeats are only occasionally observed and which provide little challenge to standard assembly efforts, HORs typically occupy multi-megabase-sized homogenized arrays that are vastly underrepresented in each chromosome assembly (IHGSC 2004;

Rudd and Willard 2004; Hayden et al. 2013). As a result, sequence descriptions of chromosome-assigned HOR arrays are currently absent from the reference assembly (Rudd and Willard 2004; Hayden et al. 2013). Characterization of satellite and non-satellite sequences that comprise a single array is necessary to initiate comparative analyses between individual genomes to study how these sequences change over time.

Individual alpha satellite arrays have been shown to vary considerably in array length and HOR sequence variants within the human population (Waye and Willard 1986; Durfy and Willard 1989; Wevrick and Willard 1989; Oakey and Tyler-Smith 1990; Warburton et al. 1991; Santos et al. 1995). This level of sequence variability is found between maternally- and paternally-inherited chromosomes (Mahtani and Willard 1990; Warburton and Willard 1995; Roizès 2006), and in whole-genome data from diploid genomes; accordingly, it is difficult to infer the long-range sequence organization and assign sequence variants to a single array. The X and Y chromosomes, represented as haploid arrays in the male genome, offer a unique opportunity for high-resolution sequence definition of individual centromeric arrays. Experimental studies of these regions have provided an initial estimate of long-range sequence organization by physical mapping (Tyler-Smith and Brown 1987; Mahtani and Willard 1990; Mahtani and Willard 1998), sampled HOR homogeneity from sequencing within the array (Durfy and Willard 1989; Schueler et al. 2001; Schindelhauer and Schwarz 2002), and demonstrated the utility of a small number of array-assigned sequence variants, or experimental markers, to study centromere evolution (Durfy and Willard 1990; Laursen et al. 1992) and population genetics within a limited number of individuals (Oakey and Tyler-Smith 1990; Santos et al. 1995; Santos et al. 2000). These centromeric surveys offer a strong foundation upon which to build comprehensive satellite DNA descriptions across the entire centromeric region within and across a large number of human genomes.

Here we model the sequence composition and local monomer organization across two alpha satellite haploid arrays on chromosomes X and Y, presenting an initial characterization of sequences currently represented by centromeric gaps in the chromosome assemblies. To accomplish this, we have designed and implemented software (linearSat) that utilizes monomer sequence and relative order as observed from a whole-genome shotgun (WGS) read database (Levy et al. 2007b), to present a linear sequence model that describes sequence content within each satellite array. To demonstrate the utility of these reference sequences we have indicated those sequences that are represented only on the X or Y chromosome, defining chromosome-specific markers useful in short-read mapping (Hayden et al. 2013). Additionally, we have provided annotation of satellite sequence copy number estimates to determine sites in the array that are low-copy and thus useful for extending long-range array characterization. Acknowledging the expected variability of satellite DNA arrays in the human population, we study this singular reference relative to 372 male, low-coverage genomes to perform a high-resolution analysis of sequence profiles and array length estimates (1000 Genomes Project Consortium 2012). This study provides an initial sequence characterization of regional centromeres from chromosomes X and Y and establishes a foundation for extending this method to study other satellite DNA arrays within complex genomes.

## RESULTS

**Algorithmic overview**

Centromeric satellite DNAs are composed of tandem repeats that, apart from a limited number of variant sites, are identical across multi-megabase-sized arrays. This excess of sequence identity and an inability to determine the correct biological ordering of repeats has challenged previous assembly algorithms. Here, we provide an alternate approach to characterizing satellite DNA arrays; abandoning the need to determine the "true" linear order, we rather aimed to generate a linear sequence that models the observed variation and repeat order as identified in an initial database of high-quality, Sanger WGS reads.

This general sequence-processing pipeline is subdivided into three steps, as depicted for the centromere X alpha satellite array in Figure 1. The HOR sequence on the X chromosome (DXZ1) is described by a highly-homogenized, 12-monomer tandem repeat that spans the length of the centromeric gap in Figure 1a (Waye and Willard 1985; Mahtani and Willard 1990). A limited number of HOR sequence variants are observed across the entirety of the array, defined by single-copy nucleotide variants (as indicated in Figure 1 by a single-base change resulting in two HORs that are 99% identical within the pink and blue boxes), rearrangements resulting in a different monomer number and organization from the canonical repeat unit (highlighted in orange), and insertion of non-satellite sequences [as shown for the long interspersed element (LINE) in green]. To study the occurrence and frequency of such variation within a given array, it is first necessary to create a read database specific to the DXZ1 array (Figure 1b, step 1). Each HOR can be defined as an ordered arrangement of alpha satellite monomers (labeled m1-m12, with average length of 171 bp) that is repeated in a head-to-tail organization [as illustrated by the red line connecting m12 back to the start of the repeat (m1)]. Due to high sequence identity among copies of a given HOR, the majority of individual monomers appear identical to the corresponding consensus monomer sequence, noted as grey ovals. However, variant sites of single nucleotide change are shown in blue, rearrangement in monomer ordering in orange (where m7 is observed adjacent to m9, omitting m8 found in the canonical organization), and sequence insertion and deletion (as illustrated in green to mark the site of LINE insertion adjacent to m10) can be readily detected within the dataset. Thus, by generating an initial description of sequence content within each centromeric array one is able to document variant sites and the relative frequency of their occurrence (Durfy and Willard 1989).

Second, the DXZ1 sequence database can be reformatted as a bidirectional multi-graph (Medvedev and Brudno 2009), as illustrated in Figure 1b, step 2. All identical full-length monomers are compressed to represent a single node, thereby emphasizing those variant sites within the array. For example, m2, containing a single-base change from the consensus, can be subdivided into two monomer groups: m2v1 (containing five identical monomers) and m2v2 (defined by one monomer). Edges between nodes are defined by local monomer adjacency and relative orientation, as observed in the initial read database. Edge weights provide the normalized frequency of observed read adjacencies from each node. Junctions with non-alpha satellite sequence, as provided for the example of the LINE element, are catalogued including the partial, interrupted alpha satellite monomer.

The final step uses a second-order Markov chain model to provide a complete traversal of the sequence graph of size N, or the estimated array length provided by sequence coverage. The algorithm is designed to include each edge at least once and in proportion to the provided edge weights. Given the homogeneity of the array, sites that differ from the consensus are expected to have low read coverage and occasionally present premature path termination. To ensure that such low-frequency variants are fully represented by the algorithm, artificial edge assignments link monomers relative to consensus-based ordering, relying on first-order information. Modeled arrangement of a centromeric array is intended to provide a more complete sequence definition of sequence variant and monomer ordering proportional to an initial unassembled read database (as shown in Figure 1b, step 3). Although it is not intended to depict the "true" long-range HOR ordering across the length of the inferred centromere array, correct long-range prediction is expected to improve as the order of the model is increased.

The resulting centromere array models provide a linearized description of the sequences within a given satellite array read database, resulting in a genomic reference useful in extending mapping tools and functional annotation.

**A comprehensive study of chromosome X and Y centromeric sequences**

To provide centromeric reference models of alpha satellite arrays on chromosomes X and Y, we prepared HOR (DXZ1 and DYZ3) read databases from a single male reference genome (HuRef) (Levy et al. 2007b)(see Methods). DXZ1 has a 12-mer HOR (Willard et al. 1983; Waye and Willard 1985), represented by 15,563 reads (totaling 13.9 Mb) and DYZ3 has a 34-mer HOR (Wolfe et al. 1985), represented by 1,008 reads (totaling 0.89 Mb). The array length estimates as determined by read depth in the HuRef genome, fall into an expected distribution of previous high-molecular weight pulse-field gel electrophoresis (PFGE) studies across a variety of cell lines for the DXZ1 array (with array lengths that vary between 1.3 and 3.7 Mb)(Mahtani and Willard 1990) and the DYZ3 array (ranging from 0.2 to 1.2 Mb)(Wevrick and Willard 1989; Oakey and Tyler-Smith 1990). In addition, we have validated PFGE array size estimates from the donor-matched cell line (data not shown), providing further support for the general findings within our study.

A study of alpha satellite sequences within each higher-order array read database provided evidence that the arrays are indeed highly homogenized (monomer global alignments to consensus; DXZ1: 97.4% average, with range 92.2-100%; DYZ3: 99.6%, with range 97.2-100%) and that interruptions in the array by non-satellite DNA are exceedingly rare (six events in the 3.6 Mb DXZ1 array, and no detectable events across the DYZ3 array). Additionally, monitoring the directionality of the monomers on both single reads and between paired reads, we find no evidence of shifts in polarity, suggesting that the HORs are organized in a single orientation across the length of the entire array in the HuRef genome. Paired-read assessment across both the X and Y arrays would suggest that the majority of paired reads contain only HOR satellite sequences, with only a small fraction (< 1.0%) assigned to uniquely mapping sequences that can be confidently assigned to p- or q-arm. Thus, these data provide evidence, in line with previous studies (Tyler-Smith and Brown 1987; Mahtani and Willard 1998), for a single alpha-satellite array that spans the length of each centromere-assigned gap for chromosomes X and Y.

To study the occurrence and frequency of array HOR sequence variants, we reformatted the read database into full-length, high-quality monomers with notation relative to the consensus monomer ordering for the DXZ1 and DYZ3 repeat units. Total monomer libraries across the repeat were consistent in both DXZ1—with an average 3,583 monomers across the 12-mer HOR—and DYZ3—with an average 58 monomers across the 34-mer HOR (SFig1a). Compression of monomers into groupings based on strict identity (edit distance 0) revealed ~10-fold compression for DXZ1 (average 300 unique monomer types) and ~20-fold compression for DYZ3 (average 3 unique monomer types) (SFig1b). Sites that vary from each HOR-derived consensus appear to be relatively equal between transitions and tranversions (as shown for DXZ1 in Figure 2), although we detect that the bases that differ from the consensus appear to deplete the total number of GC base pairs, thereby increasing AT-richness. Within the array, 8% of single-site changes are characterized by insertions or deletions relative to the consensus (insertion and deletion tracks, Figure 2). The majority (68%) of these sites are associated with the expansion or contraction of homopolymer sequences, with relatively equal changes associated with A and T nucleotides. Single- or multiple-base changes within the array provide little evidence for variation in the length of each individual alpha satellite monomer, thus maintaining the individual alpha satellite repeat unit length.

Although the majority of reads support a canonical monomer ordering, we identify ten sites of HOR rearrangement in DXZ1 (of which two have been previously described) (Willard et al. 1983; Warburton et al. 1991), (Figure 2), and three sites in DYZ3 (not shown). Investigation of the DYZ3 HOR repeat identified a low-frequency, two-monomer insertion previously determined to be represented in a smaller proportion of the array in European genomes (Wolfe et al. 1985; Santos et al. 1995). Additionally, we detect evidence for local duplication of m13, and a rearrangement involving m11 and m13 (SFig1c).

The read databases for DXZ1 and DYZ3 were reformatted into a sequence graph (as shown in Figure 1), where nodes describe a grouping of identical full-length monomers, and edges between nodes are provided based on observed local monomer ordering within a single read. In summary, we have used this data structure to describe a read database, representing the census of sequences that comprise DXZ1 and DYZ3 arrays in the HuRef genome.

**Centromere reference models for DXZ1 and DYZ3 alpha satellite arrays**

To generate a centromere reference model of each alpha satellite higher-order array, we designed freely-available software (linearSat; see Methods for link and description). This software utilizes a second-order Markov model to traverse each respective centromeric sequence graph, resulting in monomer ordering that is proportional to that observed within the initial read database. The software is sensitive to include low-coverage, variant sites within the array and employs a consensus-informed ordering that guides the extension of monomer ordering relative to the canonical repeat organization, within ambiguous regions defined by low read depth. Junctions between alpha satellite monomers and non-alpha satellite sequences described within each sequence graph are represented as those high-quality bases ($\geq$ a phred score of 20) adjacent to the full-length monomer, defined by partial alpha and non-alpha satellite sequences with an appended 100-bp gap (as first depicted in Figure 1b). Read-depth estimated array sizes for both DXZ1 and DYZ3 were used to set the threshold for Markov chain termination. In doing so, we determined the content of a 3.6-Mb DXZ1array and a 0.23-Mb DYZ3 array, representing a full listing of all monomers in the proportion expected from the input.

To evaluate the accuracy of these results, we performed a comparison of each generated array sequence to the original read database. The linearSat software operates at the level of full-length monomers, thereby omitting information from partial monomers commonly found at the 5' and 3' ends of individual sequence reads. To account for the representation of these sequences in the final projection, we reformatted the generated linear sequence and those WGS reads from the original unassembled sequence database into windows of size k (where k=50-400, with a 1-bp slide in both strands), demonstrating an average positive predictive value of 94% and 95% across all lengths k for DXZ1 and DYZ3, respectively (Figure 3a). To evaluate each inferred centromere projection for a predicted monomer ordering that is not observed in the initial read dataset, we performed inverse analysis and determined a predictive value of 84% averaged across windows in DXZ1 and 94% in DYZ3 (Figure 3b).

Given the stochastic approach of using a generative Markov process, we do not expect to generate the true long-range linear order across the entire array. However, we hypothesize that this model is capable of correctly predicting regional sequence organization (defined as greater than the length of a single read) within the resulting linear sequence. To evaluate the long-range prediction, we studied concordant paired-read support between ordered HOR within each array, demonstrating that roughly 74% of small plasmid inserts (with average insert size of 2 kb) in the DXZ1 array and 95% within the DYZ3 array have at least one concordant arrangement within the generated array. We hypothesize that longer reads, thereby increasing the model order, would greatly improve our confidence in true HOR ordering within the array. To test this, we simulated long reads from each DXZ1 and DYZ3 linearized array generated in this study and increased the model order accordingly. Blocks of correctly ordered monomers were determined against the initial DXZ1 and DYZ3 array representations. Evaluation of the maximum block length, or longest string that has an exact match with the initial array, is shown to increase with model order (within the range of 3 to 24 monomers, as shown in Figure 3c); demonstrating the ability to correctly predict the ordering of a megabase of the array (about a quarter of the estimated DXZ1 array size, 0.9 Mb) using a monomer model order of 22 (or the monomer order described on read lengths ~4 kb). Additionally, we show that extending the model order to ten (as observed within a ~2-kb read) we recover greater than 40% of the array to be correctly ordered greater than, or equal to 10 kb (SFig2a). Identifying a discrete list of monomer blocks we determined the equivalent N50 values to indicate a linear improvement with an increased model order (SFig2b). Therefore, this method is currently capable of representing local monomer ordering and sequence composition within a satellite DNA array and is expected to improve long-range organization prediction with only a modest increase in read length.

**Assessment of short read mappability across centromeric satellite arrays**

Alpha satellite DNAs are expected to share a basic sequence definition across all subsets and higher number of exact sequence alignments among closely-related HOR arrays (Alexandrov et al. 1988; Hayden et al. 2013). Such sequence homology is expected to challenge accurate mapping and interpretation of short-read datasets, common to epigenomic and population-based whole-genome sequencing studies (1000 Genomes Project Consortium 2012; ENCODE Project Consortium 2012). To establish array mappability [that is, to characterize those sequences specific to the DXZ1 and DYZ3 arrays, as described earlier (Hayden et al., 2013)], we reformatted each linearized centromeric array into a k-mer library (where k=24, 36, 50, 100-bp windows with a 1-bp slide) and identified those sequences that are found only within the DXZ1 or DYZ3 read database and lack an exact match with all remaining sequences in the HuRef genome. At the resolution of 24 bp, we detect 78% of the DXZ1 array and 49% of the DYZ3 array to be specific to those arrays, with a gain in array mappability with increasing k-mer length (SFig3). Additionally, a survey against 814 low-coverage genomes (1000 Genomes Project Consortium 2012) by sex demonstrates the specificity of DYZ3 24-mers to male individuals and a relative doubling of DXZ1 specific 24-mers when compared between females and males (SFig4), as expected. As a result, we have qualified those sites along the length of each centromeric array that are both present in the original dataset and are capable of ensuring array-specific mapping and annotation.

Sequences that comprise DXZ1 and DYZ3 can be further studied within their relative abundance within the array, thereby indicating those sites represented in the majority of HOR repeats as well as sites of low-frequency array variants. To annotate the prevalence of each k-mer library with respect to either the DXZ1 or DYZ3 sequence libraries, we estimated the frequency profiles across each linearized centromeric array. We determined 2.1% of DXZ1 24mers and 0.8% of DYZ3 24mers to be equivalent to single copy [present in original-read dataset at or below single-copy read depth estimates (see Methods) and observed less than three times in the inferred array]. Due to the expected high level of homogeneity within each array, the vast majority of sequences in both arrays are defined by consensus sequence (representing 89-92.3% of all k-mers for DXZ1 and DYZ3, respectively), with all remaining sequences representing intermediate variants that are represented only partially within either array. This initial sequence characterization—defining array specificity and sequence copy number—is intended to qualify the interpretation and mapping capabilities in these highly repetitive regions, thereby allowing these sequences to be useful as a reference for ongoing genomic studies.

**A study of centromeric array variation within human populations**

Satellite arrays are known to expand and contract through mechanisms underlying concerted evolution, resulting in substantial differences in array length and quantitative differences in higher-order repeat variants between individuals in the human population (Wevrick and Willard 1989; Mahtani and Willard 1990). To assess this level of variation in a larger set of individuals, we surveyed 372 male genomes from the 1000 Genomes Project to study sequence abundance and population-based signatures of haploid X and Y centromeric arrays (1000 Genomes Project Consortium 2012). Due to low-coverage, haploid sequence representation of X and Y arrays, we restricted our analysis to the top 75th percentile of the most frequent array-specific markers in the HuRef genome (1,546 unique 24-mers for DXZ1 and 1,837 unique 24-mers for DYZ3). To estimate pairwise similarity between individual arrays, we calculated the Euclidean distance between frequency vectors (24-mer normalized frequency profile for either DXZ1 or DYZ3 array), resulting in an n x n (372 x 372) affinity matrix. We performed unsupervised clustering (see Methods) to predict two distinct array groups (group 1 and group 2), as illustrated in heat-map clustered matrices in Figure 4a, for both DXZ1 and DYZ3. To expand our study to identify those sequence features useful in classifying each array group (i.e., capable of distinguishing DYZ3 group 1 from group 2), we performed supervised learning models (Support vector machine, SVM) with leave-one-out cross-validation (Wang et al. 2006). As a result, we identified 138 24-mers within the 90th percentile that were capable of discriminating DXZ1 groups (selected features with accuracy scores in range of 0.96 – 0.99) and 166 for DYZ3 (accuracy scores in range of 0.86 – 0.91) (STbl1, SFig5). It is expected that higher sequence coverage and longer reads will offer an increased resolution of mixing between each array group classification, as many low-frequency alleles, describing relatively new mutations that or intermediate signatures between the two groups, that are not currently included in this analysis.

To study the distribution of array lengths in the human population, we provide DXZ1 and DYZ3 size estimates for each haploid array (see Methods, SFig6), resulting in distributions that are largely concordant with previous experimental estimates from smaller numbers of individuals (Mahtani and Willard 1990; Oakey and Tyler-Smith 1990)(Table 1). The DYZ3 array is determined to have a mean size of 0.81 Mb and is observed to vary by over an order of magnitude (range, 0.1 – 2.2 Mb) (Wevrick and Willard 1989; Oakey and Tyler-Smith 1990). Likewise, we observe a mean array size of 3.2 Mb for DXZ1 (range: 0.5 – 4.9 Mb), in line with previous estimates (Mahtani and Willard 1990). When applying the previous array classification labels (group 1 and 2 for both DXZ1 and DYZ3) based on sequence signatures within each array, we identify groups of array lengths for both DXZ1 and DYZ3 that fall into two distinct bimodal distributions (t-test, DYZ3 p-value < 0.01 and DXZ1 p-value < 0.05)(Figure 4b, Table 1). Thus, we provide evidence for two predominant satellite array types in each of DXZ1 and DYZ3 that are defined by sequence composition and associated array length distribution.

To investigate population-based patterns of satellite array inheritance, we subdivided the DYZ3 groups 1 and 2 within the context of 1000 Genomes population assignments (Figure 4c). In support of previous findings (Oakey and Tyler-Smith 1990), we determine that DYZ3 group 1 arrays are observed in high frequency in Western Europeans (GBR, TSI, IBS) and that group 2 arrays are observed to be more prevalent in Asians (CHB, CHS, JPT). When monitoring population assignment of the DXZ1 array groups we observe a higher prevalence of group 1 in Asian populations (CHB, JPT, CHS), with a frequency of ~0.5 within individuals tested, and within South American populations (MXL: 0.6, CLM: ~0.5). In line with the hypothesis that the inheritance of the X and Y centromeric arrays are independent of one another, we do not observe a statistical correlation between X and Y types relative to a simulated null.

In total, these data support the hypothesis that DXZ1 and DYZ3 arrays in early human populations could be subdivided into two general groups whose genetic signatures and array size are largely maintained in modern human populations. This suggests that mechanisms of conversion and unequal crossover greatly outweigh the influence of novel mutation or inter-homologue/chromosomal exchange.

**DISCUSSION**

Addressing a long-standing technical problem for sequence assembly across genomic regions of highly homogenous repetitive DNA, here we provide an initial linear sequence using locally-ordered read assemblies of haploid human centromeric regions on the X and Y chromosomes. Within this analysis, we convert a comprehensive array sequence library into a sequence graph, thereby permitting documentation of the occurrence and frequency of sequence variants across the entirety of the array. To convert this data structure to a linear reference sequence model, we traverse a path through these centromeric sequence graphs to present repeat local ordering and array sequence variants in a manner proportional to that observed in the initial sequence read database. Further, we demonstrate by simulation the utility of this method to improve long-range ordering with a modest increase in read length. It is important to note that each linear representation provides an approximation of the true array sequence organization (as defined by the initial graph structure); however, the inferred array sequence is capable of providing a biologically rich description of array variants and local monomer organization as observed in the initial read dataset and is useful as a reference for further genomic studies. Thus, this sequence characterization and linear representation of a regional centromere address a fundamental challenge in the genome sciences—the inability to generate a reference sequence across regions of homogenized satellite DNA. This work was intended to provide a high-resolution study of two haploid arrays.

However, the method should be useful for generating a reference that represents pooled centromeric array sequence libraries from diploid chromosomes.

Genomic descriptions of human centromeric regions are necessary to promote studies of array sequence evolution and function. This requires not only single, robust array reference sequences, but also tools and annotations to guide confident and biologically meaningful alignments across highly repetitive regions. Unlike the majority of the genome that is effectively single copy in the reference assembly, satellite DNA arrays are identical across the majority of tandem repeats and, in addition, share stretches of identity with related satellite arrays distributed throughout the genome (Hayden et al. 2013). To address these issues and strengthen the utility of our reference array sequences, we identified all sites that could be localized with confidence to only the X (DXZ1) and Y (DYZ3) arrays. By thus establishing array-specific mappability, it is possible to study sequence maps within the shared HOR definition assigned to a chromosomally assigned array. The majority of these sites are shared among most, if not all, copies of the repeat within the array. To improve the resolution within each array, we have provided an index of each array-specific marker to include HOR frequency or an estimate of copy number within the HuRef genome. Such array annotation strengthens the utility of this reference database and will enable studies to extend from this singular read database to perform comparative estimates of array sequence organization within the human population.

Satellite DNA in centromeric regions had been previously shown to vary in size and proportion of HOR sequence variants within the human population (Wevrick and Willard 1989; Warburton and Willard 1992). Our evaluation of evolutionary patterns of the X and Y arrays across 372 male individuals from 14 distinct human populations (1000 Genomes Project Consortium 2012), reveals that DXZ1 and DYZ3 satellite arrays in modern humans can each be classified into one of two groups defined by sequence composition and array size. The results for DYZ3 are concordant with previous experimental estimates of two, bimodal ancestral array types identified at different frequencies within Asian and European individuals (Oakey and Tyler-Smith 1990). Here we extend that initial characterization to provide array group frequencies throughout available population-assigned genomic datasets. These data suggest that the rates of homogenization—conversion and unequal crossing over—are sufficiently high to maintain the ancestral array sequence states and sequence composition, and that introduction of novel sequences by chance mutation and/or inter-array exchange is exceedingly rare (Warburton and Willard 1995). This decrease of sequence exchange is likely expected for the Y centromere, due to the lack of homologous pairing at this site. In contrast, the DXZ1 array pairs with a homologous X chromosome and is expected to have a slightly elevated probability of sequence exchange; it therefore may be more readily influenced by molecular drive (Dover 1982; Ohta and Dover 1984). It is important to note that, due to limitations of sequence coverage, many low-frequency sequence variants capable of detecting low-proportional mixing between groups may not have been discovered in this analysis.

In total this work presents an initial, centralized centromere reference database useful for promoting additional functional and evolutionary studies to study these regions in a comparative and rigorous manner. To make these data fully accessible and integrated into current genomic studies, we have introduced an annotated reference (as shown in Figure 5) that builds upon three central results from our study: 1) the biological arrangement of repeats presented in the singular reference database; 2) mappability indexing to empower additional studies to map and further characterize these regions in an array-specific manner; and 3) a rich sequence definition across X and Y arrays in the human population. Collectively, these efforts lead to a useful genomic reference enabling studies in centromere function, satellite stability, and sequence evolution in these repetitive sites in the genome.

**METHODS**

**Alpha satellite sequence graph: DXZ1 and DYZ3**

Complete HuRef WGS read libraries for both DXZ1 and DYZ3 array were obtained from alignment to full-length HOR sequences that were previously described (Hayden et al. 2013). Alignments of DXZ1and DYZ3 HOR sequence (consensus HOR derived from sequence read library) to all remaining HuRef genomic reads (with subtraction of DXZ1 and DYZ3 reads) revealed no additional alignments across 100 consecutive bp with a threshold percent identity of 90% or greater. Reads paired to DXZ1 and DYZ3 assigned sequences were assigned to p- or q-arm using uniquely mapping read assignments to the HuRef assembly (Levy et al. 2007a)(using bwa-sw)(Li and Durbin 2010). Non-satellite repeat were identified in array-assigned reads, and high-quality paired reads using RepeatMasker (v4.0; cross-match, sensitive settings)(Smit et al. 1996-2010). Full-length monomer predictions were initially determined using hmmer software (Eddy 2009) and a model of consensus alpha satellite (Waye and Willard 1987). Limited events of spacing between monomers, where the end of one monomer does not directly precede the base of the following full-length monomer, were evaluated and corrected manually. Monomer libraries of sufficient quality (with phred score of 20 across the entirety of the full-length monomer) were organized in the same strand orientation relative to orientation of published consensus. Global alignments (EMBOSS, needle software)(Rice P 2000), provided monomer libraries relative to a reference consensus DXZ1 and DYZ3 HOR repeat unit. Sequence library viewer across DXZ1 used in Figure 2 was created using Circos software (Krzywinski et al. 2009). Monomer groups are included in the study if support for the given sequence variant is observed in donor-matched, targeted re-sequencing (Hayden et al. 2013) or observed in available flow-sorted chromosome alpha satellite sequences from X (Bentley et al. 2008) and whole chromosome flow-sorted datasets for the X (Trace Archive query terms: center_project:'CHR_X_10470') and Y chromosomes (Trace Archive query term: center_project='S228'). Identical, full-length monomers were classified under a single label (i.e., m1v1) and read descriptions were reformatted relative to the order of identical monomer groups (i.e., read id m1v1 m2v1 m3v1, etc.). Adjacency and second-order monomer information were obtained from ordered arrangements of monomer labels on each read in the array dataset.

**LinearSat algorithm**

All documentation and source code for the linearSat is freely available at github.com/JimKent/linearSat. In summary, the program requires an input of monomer chains where each vector lists monomer order as observed on sequence reads relative to a model of a given higher-order repeat and monomer variant unique identifier (e.g., Read#1 m1v1 m2v1 m3v1, etc.). Although the program is capable of generating Markov models of arbitrary order, a model order of two was used for linear representation of DXZ1 and DYZ3 arrays in this study as three full-length, high-quality monomers were observed on average per WGS Sanger read in the HuRef alpha satellite database. The program outputs a Markov model of monomer variants and sequence generated following model probabilities. When the higher-order Markov model has no data, a lower order Markov model is used. If there is no data to support a first-order Markov model, the higher-order repeat model is used instead. The probabilities within a model are reduced by the weight of the monomer output. When a variant is present only once, there is only an expectation of one copy of the variant in the output, a final pass is necessary to insert rare variants that are present in input but not output; such insertions are constrained by the higher order model.

**Intra- and inter-array mappability**

To identify those sequences that are array specific, DXZ1 and DYZ3 read databases were filtered to high-quality (where all bases within the defined k are greater than a phred score of 20) k-mer library (where k=24, 36, 48, 100 bp) in forward and reverse orientation. Reads that collectively defined each array database were subtracted from the total WGS HuRef dataset, resulting in two sequence libraries: a read library to query for specificity (either DXZ1 or DYZ3) and a remaining library of reads useful in identifying the presence and relative abundance of any given k-mer from the query as background. Each individual k-mer was considered specific to a given array if it was found in less than < 0.1% of all alpha satellite reads, and < 0.01% of all reads in the HuRef genome. To determine copy-number estimates of those k-mers that are specific to either the X or Y array, we provided raw relative frequency values (or the number of observed counts/total number of sequences within the array) and copy number estimates as derived from single copy read depth distributions across non-satellite DNA on chromosome matched datasets.

**Alpha satellite DXZ1 and DYZ3 array length estimates**

Low-coverage Illumina sequencing reads were obtained from the 1000 Genomes Project representing 372 male individuals from 14 populations (1000 Genomes Project Consortium 2012). High-quality 24-mers (phred score >20 for all bases) were compared with array specific 24-mers for DXZ1 and DYZ3. For each individual the total number of high-quality 24-mers across all reads and the total number of high-quality 24-mers exactly matching an alpha array-specific 24-mer were enumerated. For each male individual, the DYZ3 array size was first estimated as follows:

Where the proportion of all HQ 24-mers matching a DYZ3-specific 24-mer, or "a", is normalized by the proportion of all HQ 24-mers on DYZ3-containing HuRef reads matching one of the specific 24-mers, or "h", multiplied by the estimated size of the diploid male genome from hg19): $a/(h*g)$. To correct for any coverage bias, a set of 49,994 unique 24-mers was obtained from the chrY reference sequence in hg19 and enumerated across all individuals tested. These control 24-mers were matched to have the same distribution of AT-richness as the DYZ3 24-mers. To estimate the total error in the array size estimates, the size of the control region was estimated by the same calculation for each male individual and compared with the actual value. After correction to the mean coverage bias across samples, 95% of samples had control size estimates within 12% of the actual value (for DYZ3). Similarly, the DXZ1 array estimates used 49,993 unique GC-matched chromosome X 24-mers, corrected down by the mean error rate of 1.7%.

**Unsupervised clustering of array k-mer profiles**

Each array-specific satellite sequence (1,546 unique 24-mers DXZ1 and 1,837 unique 24-mers DYZ3) was surveyed for normalized abundance (total count of sequence observed normalized to the total number of high quality 24-mers within each whole genomic dataset). The resulting matrices for DXZ1 and DYZ3 (n x m), where 'n' represents each individual and 'm' provides the normalized frequency for each 24-mer queried, provided a vector profile of shared array sequence abundance with the initial HuRef reference. An affinity matrix was constructed by pairwise calculations of Euclidean distance between individual 24-mer profiles, providing a final n x n matrix of (372 individuals x 372 individuals). To provide an initial assessment of cluster number, this matrix was evaluated by hierarchical clustering displaying a heat map, or clustergram object (MATLAB, 2009b). Unsupervised, spectral clustering of two groups (Luxburg 2007) was determined by principal component analysis (MATLAB, 2009b) and consequently, K-means clustering (MATLAB, 2009b, The MathWorks; squared euclidean distance measure) to predict two clusters in both the DXZ1 and DYZ3 datasets. The optimal number of groups, or k=2, was determined as the greatest average measure of cluster proximity, or mean silhouette values (MATLAB, silhouette plot).

**Array group k-mer classification and recursive feature selection**

Similar to a previously-published method (Wang et al. 2006), individuals were classified with labels from spectral clustering (group 1 and group 2 for both DXZ1 and DYZ3 arrays). Initial matrices (n x m) were provided for both DXZ1 and DYZ3 datasets, where 'n' corresponded to each feature (or array specific 24-mer), and 'm' provided the normalized frequency (average of observed forward and reverse 24-mer) for each individual. For each feature, or 24-mer, we iteratively applied a linear Support vector machine (SVM)(R, libsvm)(Chang and Lin 2011) classifier. Leave-one-out cross-validation was used to evaluate the classifier, by which all but one individual in the dataset was used to initially train the SVM classifier to predict the class of the one held out individual. We then compare the predicted class with the actual class membership for the held out individual. Average accuracy scores across SVM cross validation were provided for each 24-mer. Features are then ranked based on the accuracy of the SVM classifier and the top 90th percentile k-mers are selected as the most informative features. Supplemental Figure 6a shows the diagram of this feature selection method.

DATA ACCESS

BioProject ID PRJNA193213; GenBank accessions: GK000058 (gi 529053718) and GK000059 (gi 529053714). Prior to the release of the GRCh38 reference, DXZ1 and DYZ3 sequence submissions have been made available from the following websites: http://hgwdev.soe.ucsc.edu/~kehayden/GenomeResearchSeq/GK000058.fa (cenX) and GK000059.fa (cenY).

All documentation and source code for linearSat is freely available at github.com/JimKent/linearSat.

ACKNOWLEDGMENTS

We thank B. Paten, A. Ewing, G. Hickey, M. Haeussler, and D. Haussler for helpful review of this work. This work was supported by grants to J.K. from NHGRI 5U41HG002371 and 3U41HG004568-09S1.

DISCLOSURE DECLARATION

The authors have nothing to declare.

FIGURE LEGENDS

Figure 1: An algorithmic overview of satellite characterization and linear representation. (a) Cartoon depiction of centromeric array spanning the complete centromere assigned gap on chromosome X. The multi-megabase-sized DXZ1 array is comprised of tandemly arranged higher-order repeats, shown as dark grey arrows. Examples of array sequence variants are indicated: between pink and blue boxes, single-nucleotide change, illustrated in the second monomer of the HOR; orange box provides a description of monomer rearrangement with a deletion in HOR monomer order; and green box demonstrating a site of transposable element insertion interrupting the repeat. (b) To generate linear representation of these sequences the algorithm uses three key steps: First, generating an array sequence database, where full length monomers that are identified on each WGS read are organized relative to the DXZ1 HOR canonical repeat, with sites of variation as indicated. Second, read databases are reformatted into sequence graphs, wherein nodes are defined by identical monomers and edge weights are defined by the normalized read counts that define each observed adjacency in the WGS read database. Finally, traversal of the graph using a second-order Markov model provides a linear description of the original read database: presenting variant sequences in the proportion and preserving the local-monomer ordering (defined by length of read database ~500bp) as observed in the initial read database.

Figure 2: A complete array sequence database across centromeric regions. Monomer sequence identity across each monomer with average percent identity across a 10-bp window, with red color increasing to 100% as provided in key. Transitions (green) and transversions (blue) relative to the consensus sequence are provided for each 10-bp window (where the sum of each paired transition frequency window and transversion frequency window is 1). Sites of single base-pair insertion (white tracks with dark grey background) and deletion (dark grey on light grey background) are provided as observed in monomer library. Junctions that describe insertions of RepeatMasker-identified transposable elements are shown in purple with numbers indicating read depth. Consensus links (>3000 read support) between individual monomers are shown in black, non-consensus links describing rearrangements in the HOR repeat structure ordering are shown in shades of blue, with color intensity increase with estimated copy number. Image was created using the Circos software (Krzywinski et al. 2009).

Figure 3: Evaluation of linear representation of centromeric arrays. (a) Estimate of accurate WGS sequences in processed linear representation of X (black) and Y (grey) linearized centromeric arrays. Read libraries and linearized centromere arrays X and Y are reformatted into k-mer libraries (where k=50-400 bp with 1-bp slide in both strand orientations) and the proportion of sequences observed in the initial read database observed in the final database. (b) Estimate of sequences observed in linearized centromeric array relative to the initial WGS sequence database, where proportions less than one reflect the gain of novel sequence windows due to the Markov chain model. (c) To determine the improvement of array long-range prediction given an increase of model order, simulated long reads were generated at random from each linearized centromeric array (with length defined by monomer order 3-23, with an average monomer of 171 bp) and the longest arrangement of correctly ordered monomers were normalized to the total length of the array.

Figure 4: Assessment of array variation in the human population. (a) Hierarchical clustering and heatmap representation of affinity matrices for array-specific 24-mer frequencies across the X and Y centromeres provide evidence for two array groups (1 and 2). (b) Classification labels from spectral clustering of array 24-mer profiles for each individual array demonstrate a bimodal distribution with observed array size (DYZ3 group 1 in blue, group 2 in red; DXZ1 group1 in yellow, group 2 in purple). Population-based labels assign array groups to particular geographic locations (c).

Figure 5: Centromeric reference database and sequence annotation. Linear representation of the DYZ3 array is shown to completely replace the centromere gap placeholder in the chromosome Y reference assembly. Evaluation of monomer ordering across the array predicts 40 higher-order repeat units within a generated array of 227 kb. Increased resolution in the linearized centromeric array demonstrates the monomer sequence order along the bottom in blue shading (labels m1v-m34v) that defines the particular HOR arrangement, and the variant sites and base changes observed in the dataset (shown in purple). Each 24-bp sliding window across this region demonstrates the representation of these sequences within the HuRef WGS database, with peaks indicating sites that are overrepresented and likely due to exact homology with satellites outside of the Y array. The top 75[th] percentile mappable sites are provided to extend survey across other individuals. Six individual array profiles are provided as an example of population-based data, where DYZ3 array group 1 (three individuals from the CEU population) is shown in blue, and array group 2 (three individuals from the CHS population) is shown in red.

Table1: Array length estimate of DXZ1 and DYZ3.

| Chr (locus) | Array Group | No. of Arrays | Mean Size (Mb) | Range (Mb) |
|---|---|---|---|---|
| X (DXZ1) | 1 | 242 | 2.698 | 0.705–4.166 |
| X (DXZ1) | 2 | 130 | 4.062 | 3.043–8.313 |
| Y (DYZ3) | 1 | 181 | 0.399 | 0.126–0.823 |
| Y (DYZ3) | 2 | 191 | 1.186 | 0.774–2.389 |

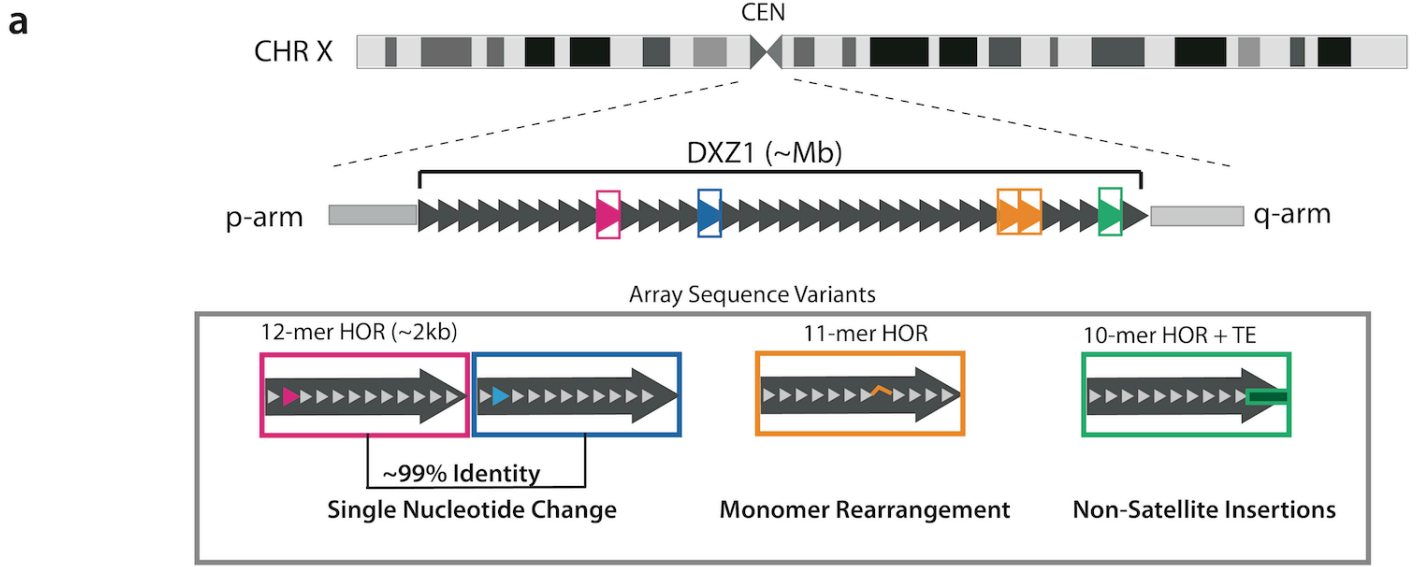

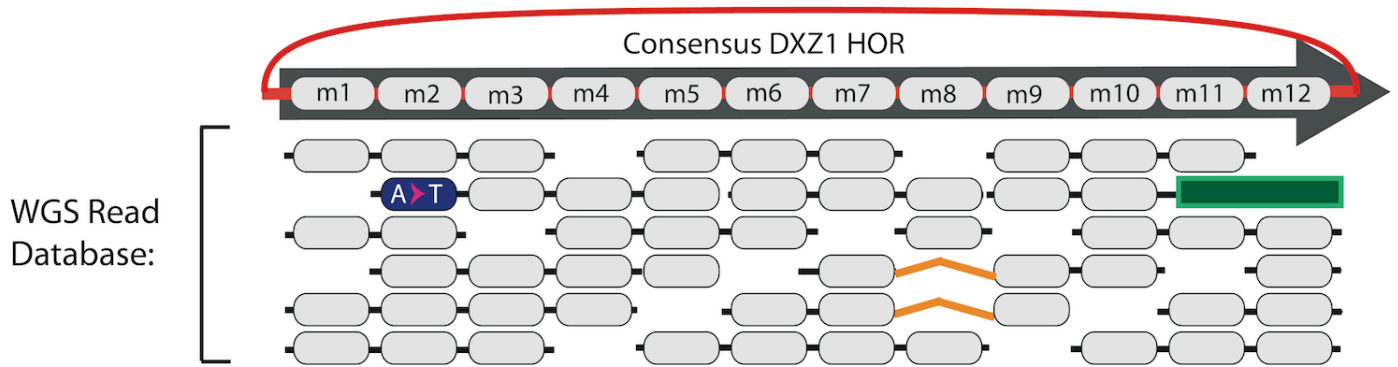

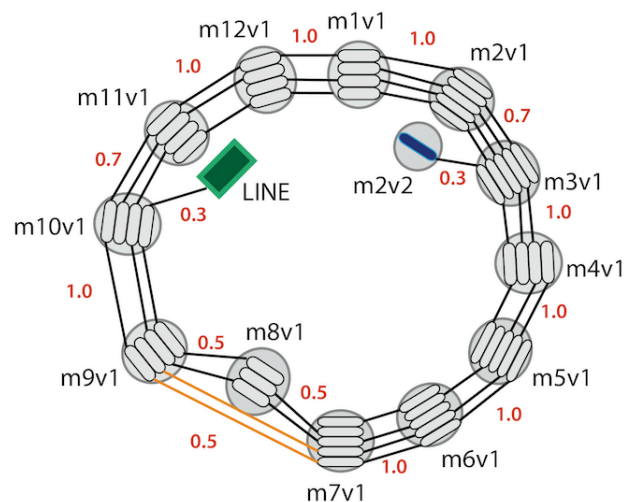

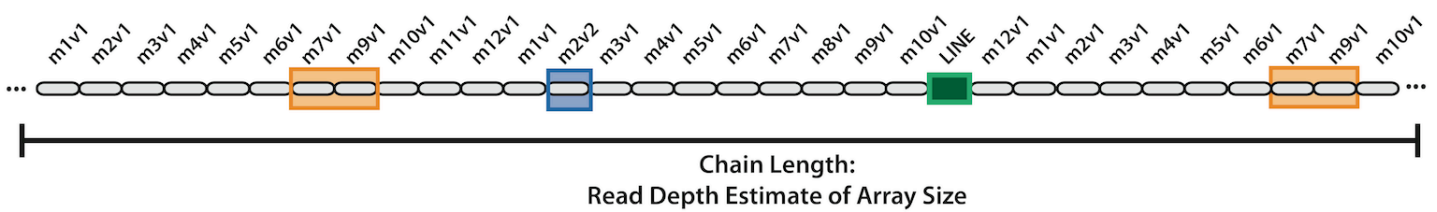

Figure 2

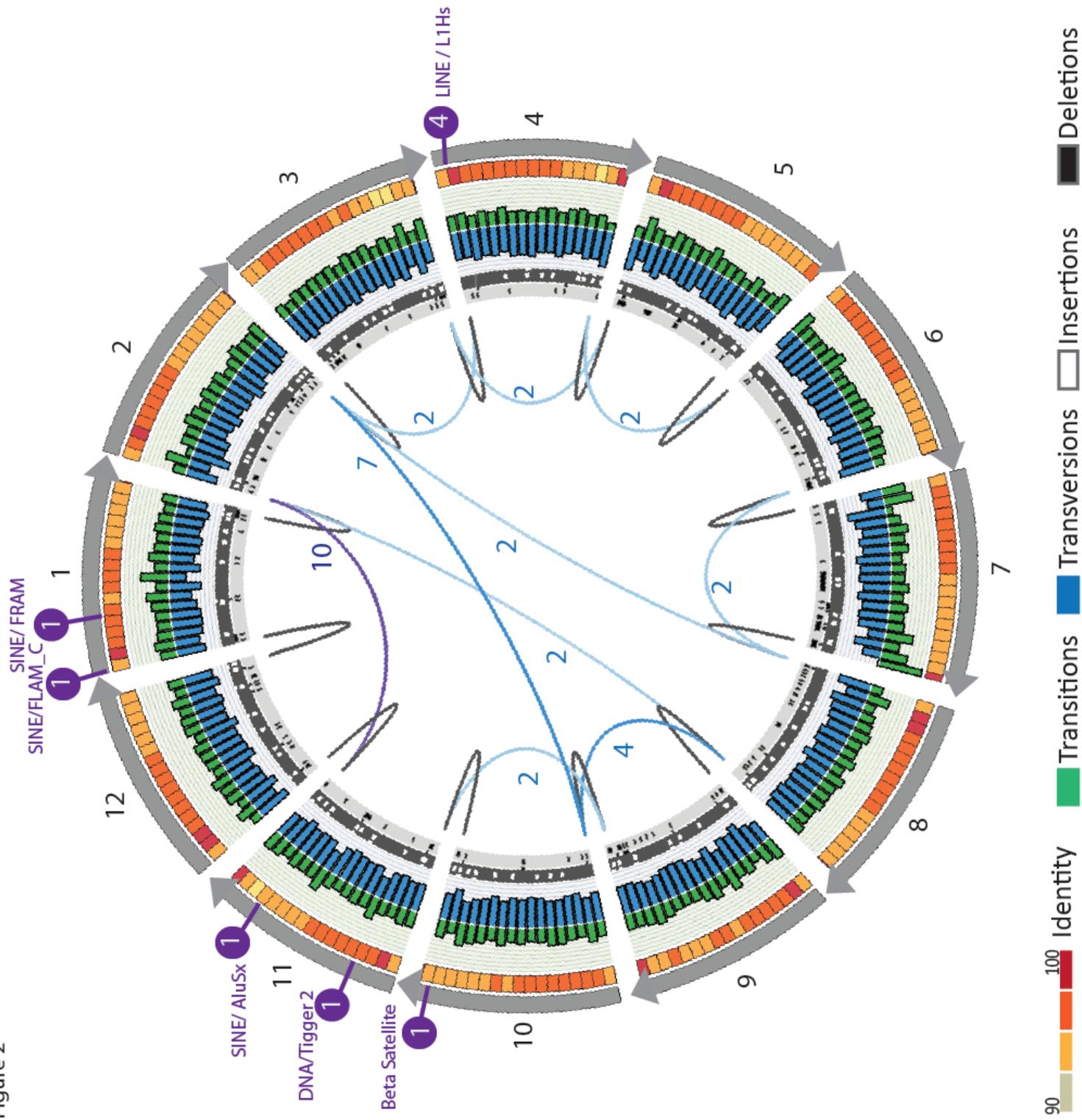

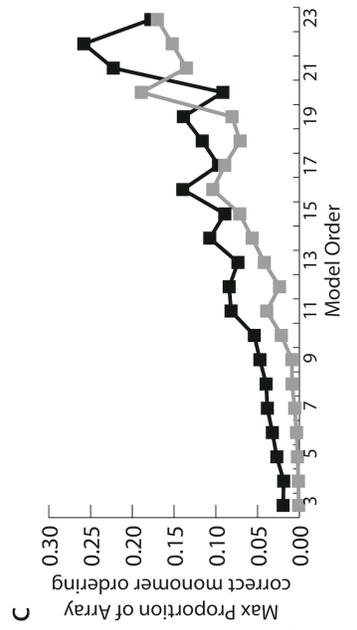

a

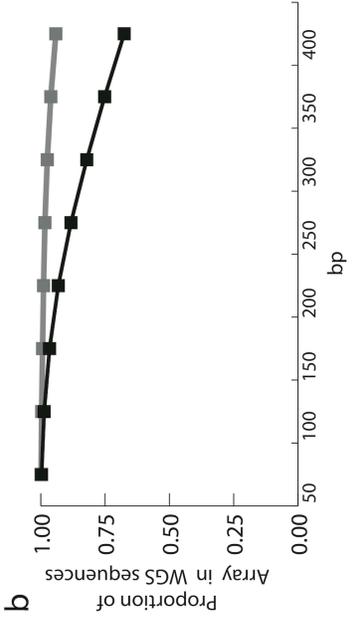

b

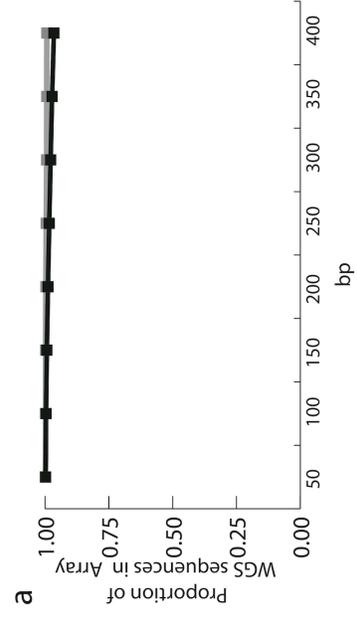

c

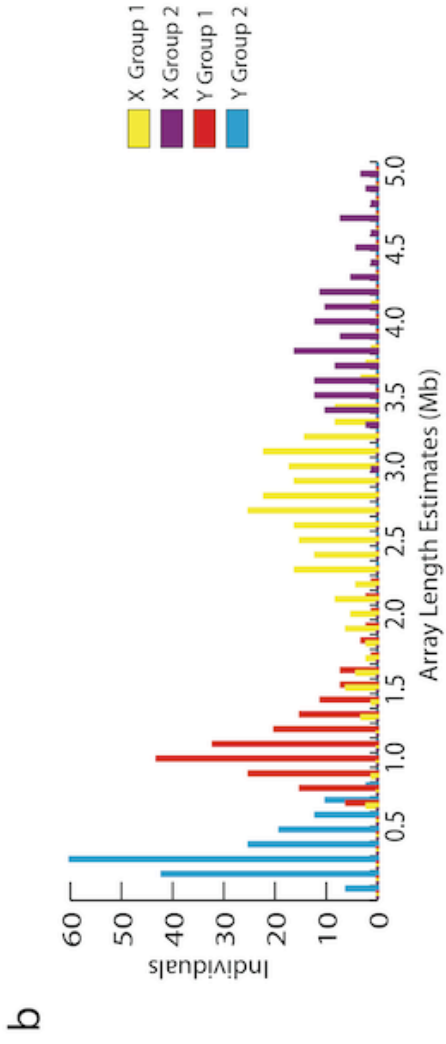
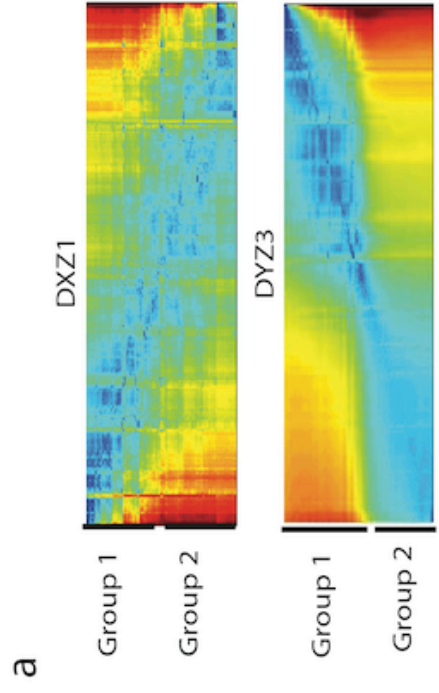
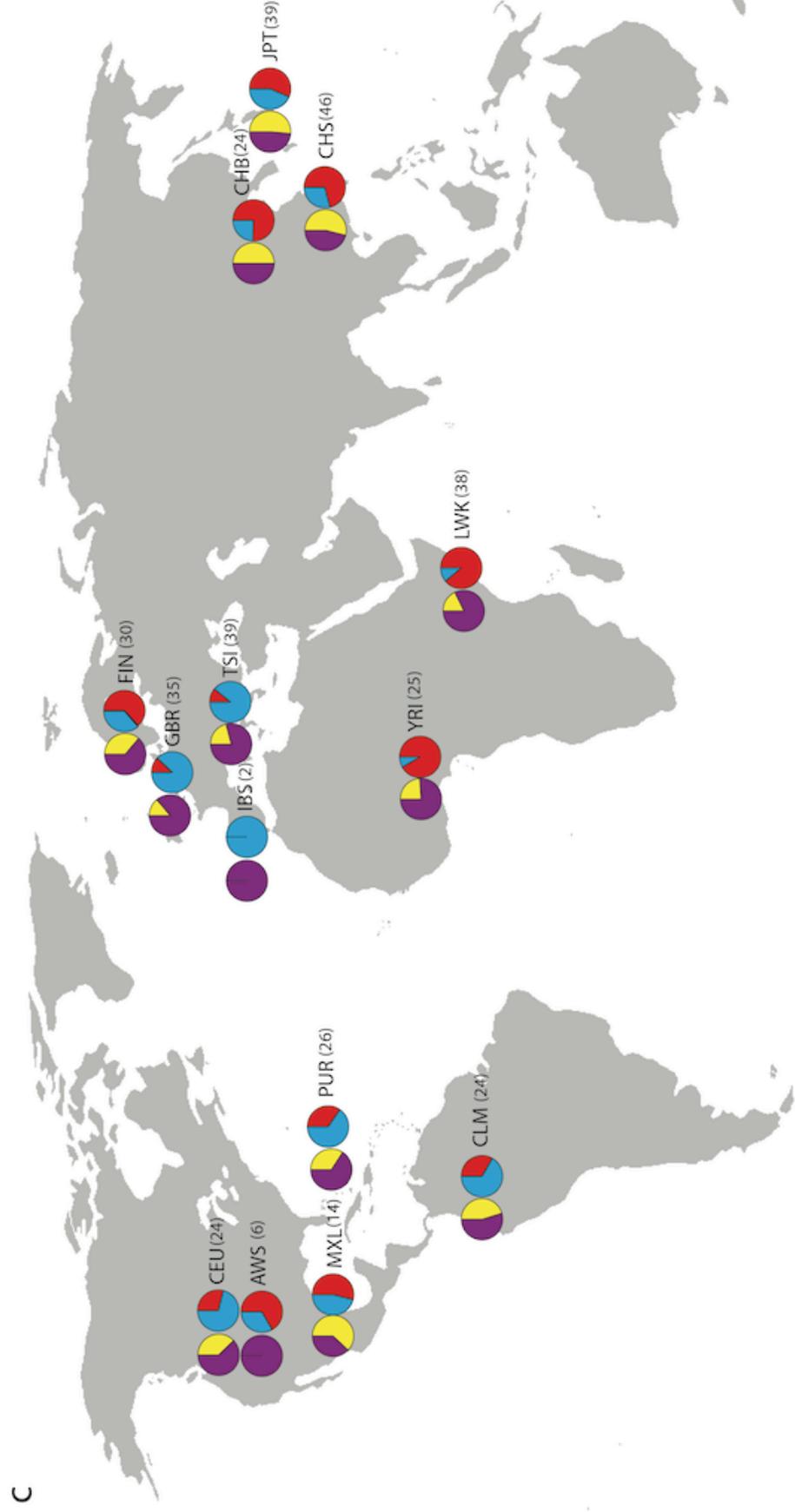

chrY

DYZ3 HOR ARRAY

Y Array Group 1 Haplogroup R: HG00103, HG00108, HG00112

Y Array Group 2 Haplogroup O: HG00650, HG00653, HG00656

24mer Mappable
24mer Profile
Array Variants: C/G, T/A, C/T, C/A
DYZ3 HOR
Monomer Components: m1v0, m2v1, m3v0, m4v2, m5v0, m6v0, m7v0, m8v1, m9v0, m10v0, m11v0, m12v0, m13v0, m14v0, m15v0, m16v0, m17v1, m18v0, m19v0, m20v0, m21v0, m22v1, m23v0, m24v0, m25v0, m26v0, m27v0, m28v0, m29v0, m30v0, m31v0, m32v0, m33v0, m34v0

# Supplemental Figures and Table Legends

## Centromere reference models for human chromosomes X and Y satellite arrays

**SHORT TITLE:** Linear sequence models of human centromeric DNA

Karen H. Miga[1,2], Yulia Newton[1], Miten Jain[1], Nicolas F. Altemose[2,3], Huntington F. Willard[2] and W. James Kent[1]*

[1] Center for Biomolecular Science & Engineering, University of California Santa Cruz, Santa Cruz, California, United States of America

[2] Duke Institute for Genome Sciences & Policy, Duke University, Durham, North Carolina, United States of America.

[3] Department of Statistics, University of Oxford, Oxford, United Kingdom
.

*Address for Correspondence:
W. James Kent
(831) 459-1401
Center for Biomolecular Science & Engineering
University of California
MS: CBSE-ITI
1156 High Street
Santa Cruz, CA 95064
kent@soe.ucsc.edu

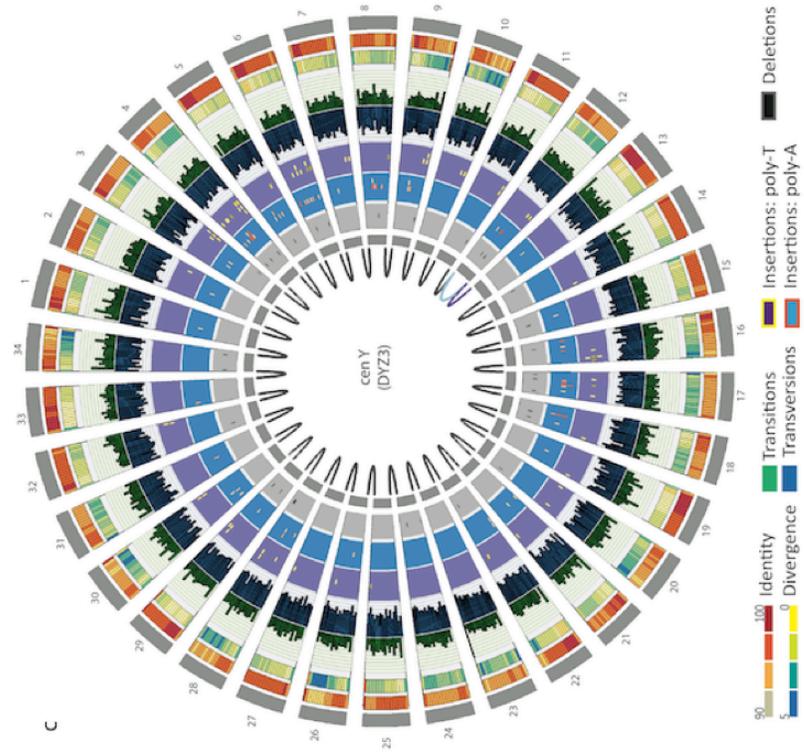

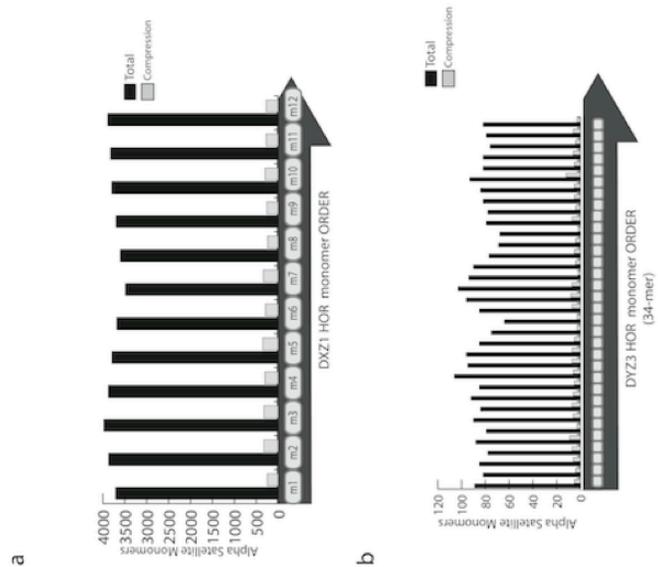

**Supplemental Figure 1:** Monomer depth across the array arranged relative to the DXZ1 12-mer canonical repeat (a) and DYZ3 34-mer canonical repeat (b) and the compression based on identical monomer groups demonstrate a ~10-fold and ~20-fold compression, respectively. (c) A complete array sequence database across chromosome Y centromeric array (DYZ3, 34-mer). Monomer sequence identity across each monomer with average percent identity across a 10-bp window, with red color increasing to 100% as provided in key for sequence identity (conversely, divergence is shown 5-0% in heat map blue to yellow). Transitions (green) and transversions (blue) relative to the consensus sequence are provided for each 10-bp window (where the sum of each paired transition frequency window and transversion frequency window is 1). Sites of poly-A (blue background with orange tracks) and poly-T (purple background with yellow tracks), with deletion (light grey background with dark grey tracks) are provided as observed in monomer library. Consensus links (>40 read support) between individual monomers are shown in black, non-consensus links describing rearrangements in the HOR repeat structure ordering are shown in shades of blue, with color intensity increase with estimated copy number. Image was created using the Circos software (Krzywinski et al. 2009).

a

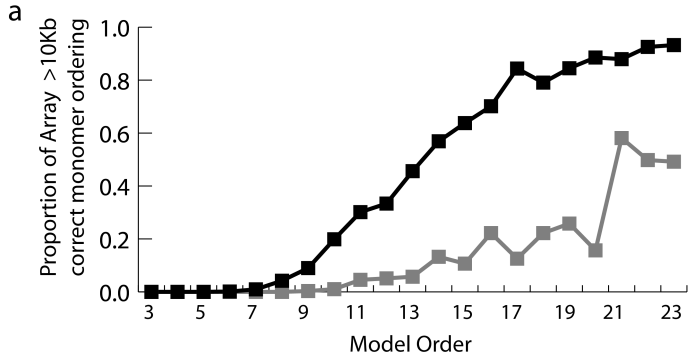

b

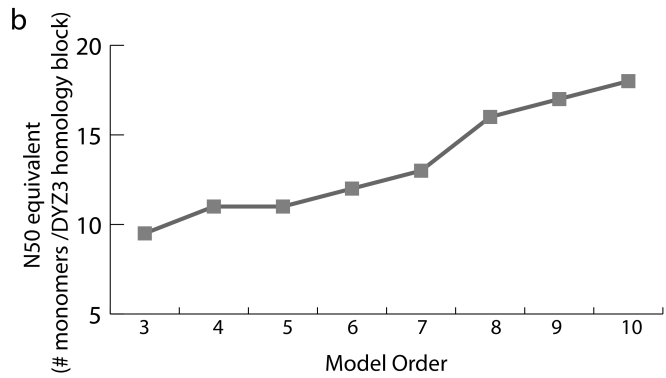

**Supplemental Figure 2:** Improvement of long-range array sequence ordering using simulated read data from linearized array DXZ1 (black) and DYZ3 (grey). (a) Increasing monomer order (3 monomers to 23 monomers) provides an increase in the number of windows that are 10 kb or greater that exactly match the initial DXZ1 or DYZ3 array. (b) Similar to an N50 evaluation, we subdivided the generated arrays (using model order 3–10) for the DYZ3 array into discrete blocks of the longest stretch of homology, the blocks were rank ordered by size and the median block provided (listed as the number of monomers correctly ordered).

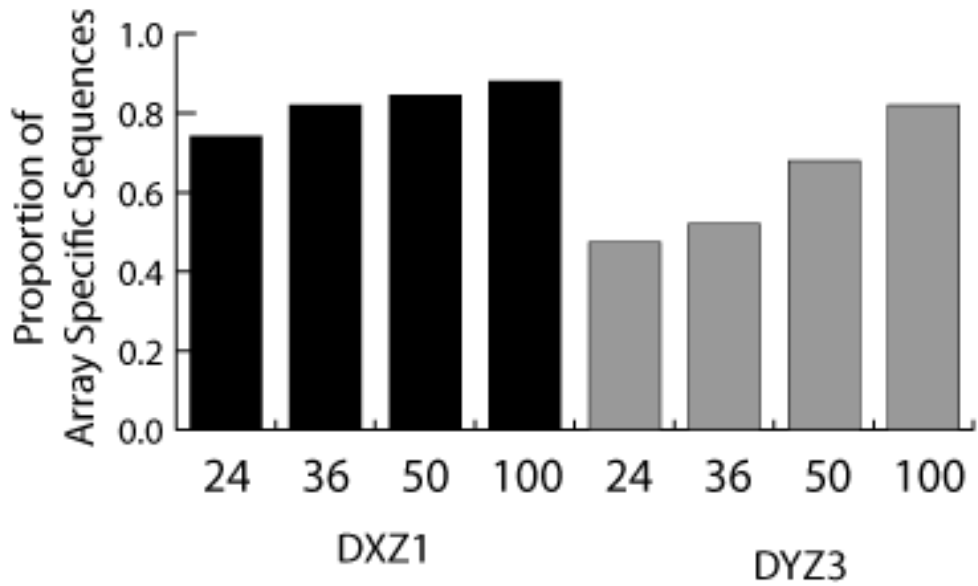

**Supplemental Figure 3:** Read libraries from DXZ1 and DYZ3 arrays were subdivided into windows of length k (where k=24, 36, 50, 100 bp). The proportion of each k-mer library that is array specific relative to all other 24mers in the HuRef genome are shown to increase with window size and vary based on the sequence definition of a given array (DXZ1 versus DYZ3).

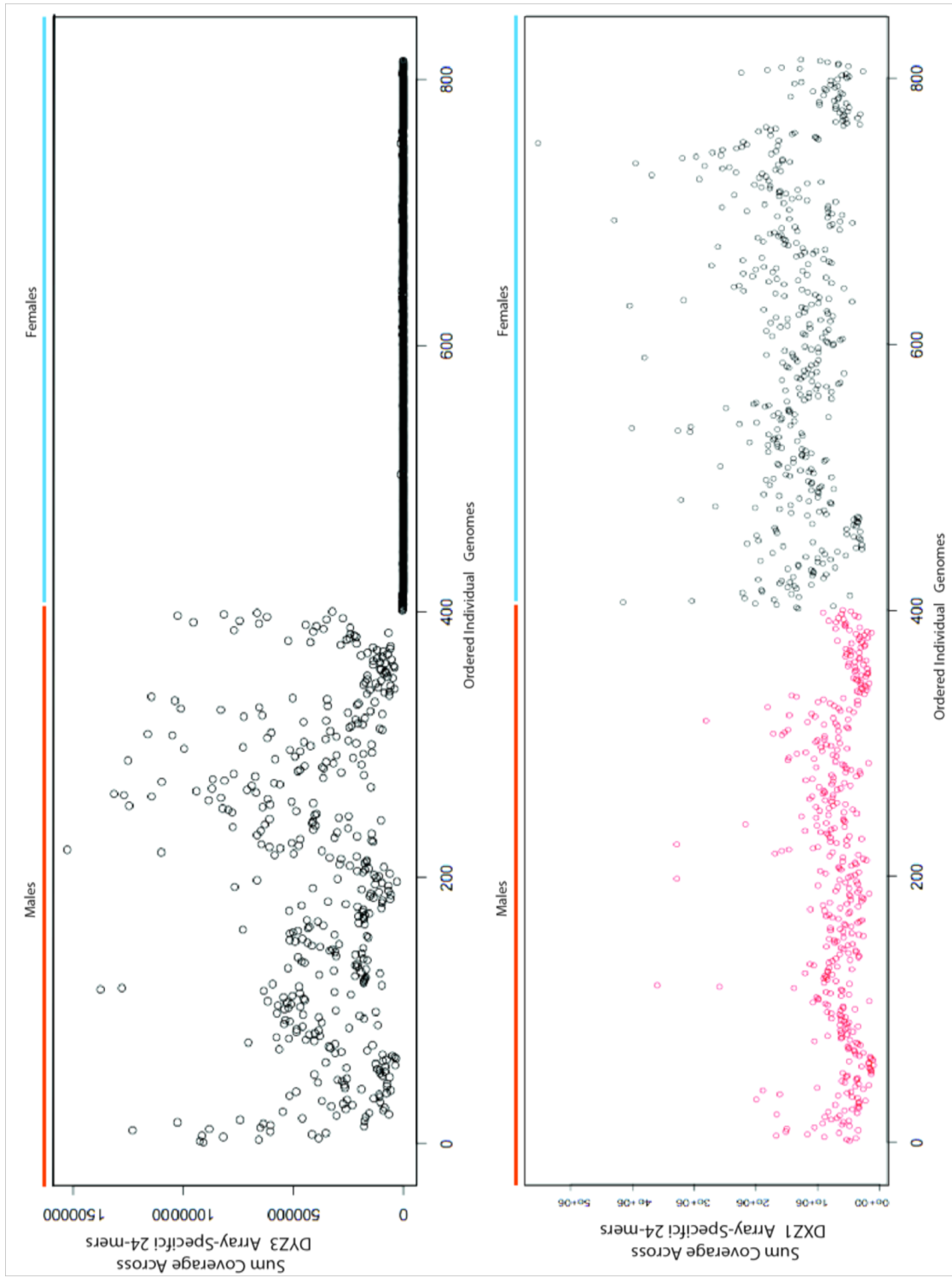

**Supplemental Figure 4:** Sequence coverage summary across 24-mer array specific libraries DYZ3 (1839 24-mers) and DXZ1 (1546 24-mers) in 814 individuals (low-coverage whole genome datasets). Individuals are sorted by gender, with males indicated in red and females indicated in blue.

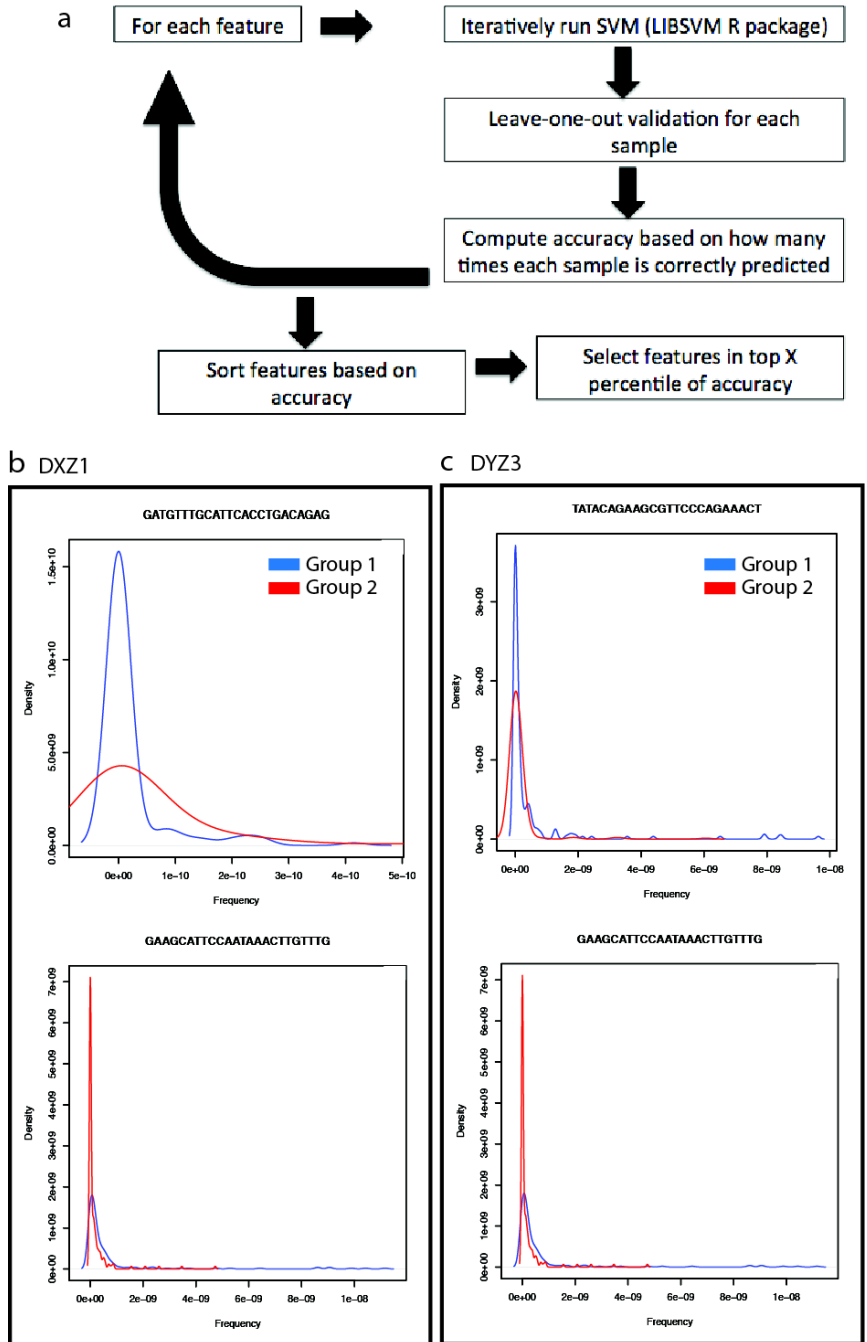

**Supplemental Figure 5:** Feature selection to identify group informative 24-mers **(a)** A flow-diagram of SVM leave-one out cross validation feature selection method employed to identify those 24-mers capable of discriminating array group1 from group 2, in both DXZ1 and DYZ3 arrays. Features were selected in the top 90th percentile of accuracy, with two examples given for group1 and group2 for both arrays (DXZ1, b; DYZ3, c). For each example 24-mer, individuals from group 1 are shown in blue and group 2 in red, with the genome-wide frequency, as observed across each individual in their respective group.

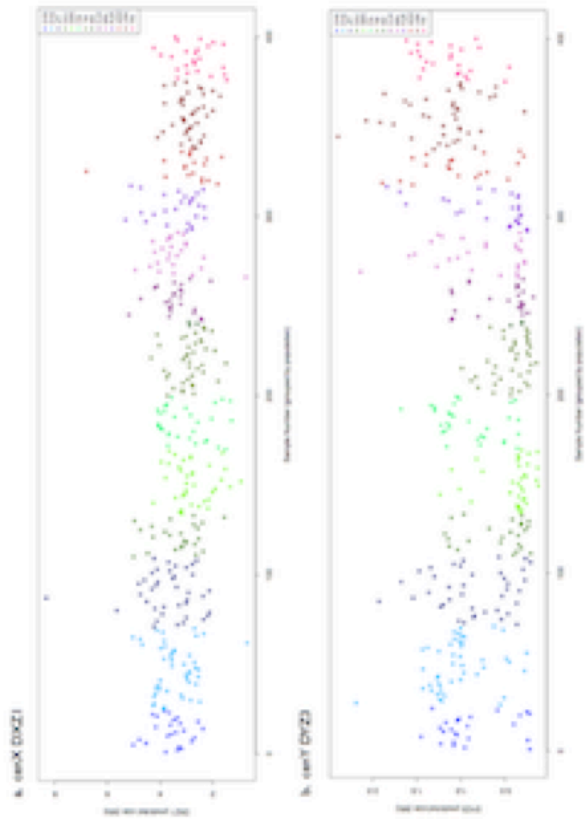

**Supplemental Figure 6:** Array length estimates for DYZ3 and DXZ1. On top, DXZ1 and array length estimates (Mb) for each population classification with population labels (as shown in color key). On bottom, DYZ3 array size estimate (Mb) for population assignments (referencing population color labels).

**Supplemental Table 1:** A sequence description of 24-mers identified in the 90th percentile SVM leave one out cross validation classifier for groups 1 and 2. Column 1: unique identifier (uid), column 2: provides array label of DXZ1 (chrX) or DYZ3 (chrY), column 3 and 4 provide each selected 24-mer in forward and reverse orientation, and column 5 provides a list of observed restriction sites.

Krzywinski M, Schein J, Birol I, Connors J, Gascoyne R, Horsman D, Jones SJ, Marra MA. 2009. Circos: an information aesthetic for comparative genomics. *Genome Res* **19**(9): 1639-1645.

Supplemental Table 1: 24mers identified in 90th Percentile SVM (leave one out validation).

| uid | Array | 24-mer F | 24-mer R | Restriction Sites |
|---|---|---|---|---|
| 0 | dxz1 | GTTCAAAACACTCTTTCTGTAGAA | TTCTACAGAAAGAGTGTTTTGAAC | AgsI,MspJI,SfeI*,MspJI |
| 1 | dxz1 | GCCTACGGTAGTACAGGAAGGAAG | CTTCCTTCCTTCCTGTACTACCGGC | MspJI,Tsp4CI*,TatI,Csp6I,Hin4II* |
| 2 | dxz1 | CTCAGAAACTTCTGAGTGATGAAT | ATTCATCACTCAGAAGTTTCTGAG | MspJI,DdeI,BspCNI,BseMII,Hpy188I,MspJI,Hpy188I,DdeI,BspCNI,BseMII |
| 3 | dxz1 | GGTTGAACCCTCCTTTTGATTGAG | CTCAATCAAAAGGAGGGTTCAACC | AgsI,MnlI,MspJI,MspJI |
| 4 | dxz1 | CTTTCGAAACGGGAATATTTCCAC | GTGGAAATATTCCCGTTTCGAAAG | AsuII,TaqI,MspJI,SspI |
| 5 | dxz1 | CTTTTGAAAGAGCAGCTATAAAAC | GTTTTATAGCTGCTCTTTCAAAAG | AgsI,BlsI,BisI,TseI,SetI,CviJI,AluI,FaiI |
| 6 | dxz1 | CATAACTAAGCACAAACACTCTGA | TCAGAGTGTTTGTGCTTAGTTATG | MspJI,FaiI,MspJI,DdeI,Hpy188I,MspJI |
| 7 | dxz1 | AACTCCCAGGGTTGAACATTCCTT | AAGGAATGTTCAACCCTGGGAGTT | PasI,MspJI,SecI*,MspJI,LpnPI,BssKI,SecI*,ScrFI,BciT130I,EcoRII,MspJI,AgsI |
| 8 | dxz1 | TGTATTCAGCTCCCAGAGTTGAAC | GTTCAACTCTGGGAGCTGAATACA | SetI,CviJI,AluI,AgsI |
| 9 | dxz1 | ACGGTAGTACAGGAAGGAAGTTCA | TGAACTTCCTTCCTGTACTACCGT | Tsp4CI*,TatI,Csp6I,RsaI,Hin4II* |
| 10 | dxz1 | GTGATGACGGAGTTTCACTCACAG | CTGTGAGTGAAACTCCGTCATCAC | TspGWI |
| 11 | dxz1 | TATTTGGACTTCTTTGAAGGTTTC | GAAACCTTCAAAGAAGTCCAAATA | AgsI,SetI,Hin4II* |
| 12 | dxz1 | CCTTCATTTGAAACGTCTATATCT | AGATATAGACGTTTCAAATGAAGG | Hin4II*,AgsI,TaiI,SetI,MaeII,FaiI |
| 13 | dxz1 | AATGGAGATTTGCACTGCTTTGAG | CTCAAAGCAGTGCAAATCTCCATT | CviRI*,TspRI,BtsIMutI,BtsI,MspJI,MspJI |
| 14 | dxz1 | AGTGGTAGTGAAGGAAAGAACTTC | GAAGTTCTTTCCTTCACTACCACT | Hin4II* |
| 15 | dxz1 | GAGTTCAGGTTCAAAACACTCTTT | AAAAGAGTGTTTTGAACCTGAACTC | MspJI,SetI,AgsI |
| 16 | dxz1 | ACACGTTTTGCAGAATCTGCAAGG | CCTTGCAGATTCTGCAAAACGTGT | AflIII,MspJI,TaiI,SetI,MaeII,CviRI*,MwoI,BstAPI,AlwNI,TfiI,HinfI,CviRI*,MspJI,MspJI |
| 17 | dxz1 | TATTCTTTGTGATGACGGAGTTTC | GAAACTCCGTCATCACAAAGAATA | |
| 18 | dxz1 | ATGGAGATTTGCACTGCTTTGAGG | CCTCAAAGCAGTGCAAATCTCCAT | CviRI*,TspRI,BtsIMutI,BtsI,MspJI,MspJI,MnlI |
| 19 | dxz1 | TCTTTCGAAACGGGAATATTTCCA | TGGAAATATTCCCGTTTCGAAAGA | AsuII,TaqI,MspJI,SspI |
| 20 | dxz1 | GCATTCACCTGACAGAGTTGAACT | AGTTCAACTCTGTCAGGTGAATGC | SetI,AgsI |
| 21 | dxz1 | TCTGAGTGATGAATGCATTCAAGT | ACTTGAATGCATTCATCACTCAGA | Hpy188I,MspJI,DdeI,BsmI,XmnI,EcoT22I,CviRI*,AgsI,BsmI,MspJI |
| 22 | dxz1 | ATTCAACTCCCAGGGTTGAACATT | AATGTTCAACCCTGGGAGTTGAAT | AgsI,PasI,SecI*,BssKI,SecI*,ScrFI,BciT130I,EcoRII,AgsI |
| 23 | dxz1 | TTTCGTTGGAAAAGGGAATAATTT | AAATTATTCCCTTTTCCAACGAAA | TspEI* |
| 24 | dxz1 | GTTGGAAAAGGGAATAATTTCCCA | TGGGAAATTATTCCCTTTTCCAAC | XmnI,TspEI* |
| 25 | dxz1 | ACCTCTCAGAGTTGAACTTTCCCT | AGGGAAAGTTCAACTCTGAGAGGT | MnlI,MspJI,DdeI,BspCNI,BseMII,Hpy188I,MspJI,AgsI |
| 26 | dxz1 | GATGAATGCATTCAAGTCACACAG | CTGTGTGACTTGAATGCATTCATC | TspDTI,BsmI,XmnI,EcoT22I,CviRI*,AgsI,BsmI,Tsp45I,MaeIII |
| 27 | dxz1 | GTGATGAATGCATTCAAGTCACAC | GTGTGACTTGAATGCATTCATCAC | TspDTI,BsmI,XmnI,EcoT22I,CviRI*,AgsI,BsmI,Tsp45I,MspJI,MaeIII |
| 28 | dxz1 | GGTGGATATTTGGACCACTCTGAG | CTCAGAGTGGTCCAAATATCCACC | AvaII,AsuI*,Hpy188I,DdeI,MspJI,MspJI,BspCNI,BseMII |
| 29 | dxz1 | CCGCAAGGGGATATTTGGACTTCT | AGAAGTCCAAATATCCCCTTGCGG | AciI,MspJI,MspJI,MspJI |
| 30 | dxz1 | CGGTAGTACAGGAAGGAAGTTCAT | ATGAACTTCCTTCCTGTACTACCG | TatI,Csp6I,RsaI,Hin4II*,TspDTI |
| 31 | dxz1 | TCATGATGAATGCATTTAACTTGC | GCAAGTTAAATGCATTCATCATGA | Hpy178III*,BspHI,MspJI,FatI,FaiI,NlaIII,CviAII,TspDTI,BsmI,EcoT22I,CviRI*,MseI,MspJI |
| 32 | dxz1 | CTTCGTTTGAAACATCTATATCTT | AAGATATAGATGTTTCAAACGAAG | AgsI,FaiI |
| 33 | dxz1 | CATTCAACTCCCAGGGTTGAACAT | ATGTTCAACCCTGGGAGTTGAATG | AgsI,PasI,SecI*,BssKI,SecI*,ScrFI,BciT130I,EcoRII,AgsI |
| 34 | dxz1 | TTGAAACATCTATATCTTCACATC | GATGTGAAGATATAGATGTTTCAA | AgsI,FaiI,MboII |
| 35 | dxz1 | CTAAAAGCTAAACGGAAGCACTCT | AGAGTGCTTCCGTTTAGCTTTTAG | MspJI,SetI,CviJI,AluI |
| 36 | dxz1 | CACCTCTCAGAGTTGAACTTTCCC | GGGAAAGTTCAACTCTGAGAGGTG | SetI,MnlI,MspJI,DdeI,BspCNI,BseMII,Hpy188I,AgsI |
| 37 | dxz1 | TCATTTGAAACGTCTATATCTTCA | TGAAGATATAGACGTTTCAAATGA | AgsI,TaiI,SetI,MaeII,FaiI,MboII |
| 38 | dxz1 | TTTGCATTCACCTGACAGAGTTGA | TCAACTCTGTCAGGTGAATGCAAA | CviRI*,BsmI,SetI |
| 39 | dxz1 | TTCATGATGAATGCATTGAACTCG | CGAGTTCAATGCATTCATCATGAA | Hpy178III*,BspHI,MspJI,FatI,FaiI,NlaIII,CviAII,TspDTI,BsmI,EcoT22I,CviRI*,AgsI,MspJI |
| 40 | dxz1 | TGACAGAGTTGAACTTTCCCTTTG | CAAAGGGAAAGTTCAACTCTGTCA | MspJI,AgsI,MspJI |
| 41 | dxz1 | ATTTCTTTCGAAACGGGAATATTT | AAATATTCCCGTTTCGAAAGAAAT | AsuII,TaqI,SspI |
| 42 | dxz1 | TAAGCACAAACACTCTGAGAAAGT | ACTTTCTCAGAGTGTTTGTGCTTA | MspJI,MspJI,Hpy188I,DdeI,MspJI,BspCNI,BseMII |
| 43 | dxz1 | TAAACTGAAGCATTCTCGGAAACT | AGTTTCCGAGAATGCTTCAGTTTA | MspJI,BsmI,Hpy188I,MspJI,MspJI |
| 44 | dxz1 | AAAATGGAGATTTGCACTGCTTTGA | TCAAAGCAGTGCAAATCTCCATTT | CviRI*,TspRI,BtsIMutI,BtsI,MspJI |
| 45 | dxz1 | TTCTTTCGAAACGGGAATATTTCC | GGAAATATTCCCGTTTCGAAAGAA | AsuII,TaqI,MspJI,SspI |
| 46 | dxz1 | GACAGAGTTGAACTTTCCCTTTGA | TCAAAGGGAAAGTTCAACTCTGTC | MspJI,AgsI,MspJI |
| 47 | dxz1 | ATGTTTGCATTCACCTGACAGAGT | ACTCTGTCAGGTGAATGCAAACAT | CviRI*,BsmI,SetI,HphI |
| 48 | dxz1 | TTGCACTGCTTTGAGGCCTACGGT | ACCGTAGGCCTCAAAGCAGTGCAA | CviRI*,TspRI,BtsIMutI,BtsI,StuI,CviJI,HaeIII,MnlI,Tsp4CI*,MspJI |
| 49 | dxz1 | TGAATGCATTTAACTTGCAGAGAT | ATCTCTGCAAGTTAAATGCATTCA | BsmI,EcoT22I,CviRI*,MseI,CviRI* |
| 50 | dxz1 | TGTTTGTATTCAGCTCCCAGAGTT | AACTCTGGGAGCTGAATACAAACA | SetI,CviJI,AluI,MspJI |
| 51 | dxz1 | TACGGTAGTACAGGAAGGAAGTTC | GAACTTCCTTCCTGTACTACCGTA | Tsp4CI*,TatI,Csp6I,RsaI,Hin4II* |
| 52 | dxz1 | AGCTATGAAACACTCCTTTTCGAG | CTCGAAAAGGAGTGTTTCATAGCT | CviJI,AluI,FaiI,TspDTI,TaqI,MspJI,MspJI |
| 53 | dxz1 | ATTGCATTCAACTCCCAGGGTTGA | TCAACCCTGGGAGTTGAATGCAAT | CviRI*,AgsI,BsmI,PasI,SecI*,BssKI,SecI*,ScrFI,BciT130I,EcoRII,MspJI,MspJI,LpnPI |
| 54 | dxz1 | GCCTTCATTTGAAACGTCTATATC | GATATAGACGTTTCAAATGAAGGC | Hin4II*,AgsI,TaiI,SetI,MaeII,FaiI |
| 55 | dxz1 | CAAATGGAGATTTGCACTGCTTTG | CAAAGCAGTGCAAATCTCCATTTG | MspJI,CviRI*,TspRI,BtsIMutI,BtsI,MspJI |
| 56 | dxz1 | TGTGATGTTTGTATTCAACTGCCA | TGGCAGTTGAATACAAACATCACA | AgsI |
| 57 | dxz1 | GAATCTGCAAATGGAGATTTGCAC | GTGCAAATCTCCATTTGCAGATTC | TfiI,HinfI,CviRI*,CviRI*,MspJI |
| 58 | dxz1 | AAGCACAAACACTCTGAGAAAGTT | AACTTTCTCAGAGTGTTTGTGCTT | MspJI,MspJI,Hpy188I,DdeI,MspJI,BspCNI,BseMII |
| 59 | dxz1 | TAACTAAGCACAAACACTCTGAGA | TCTCAGAGTGTTTGTGCTTAGTTA | MspJI,DdeI,Hpy188I,DdeI,MspJI,MspJI,BspCNI,BseMII |
| 60 | dxz1 | ACTAAGCACAAACACTCTGAGAAA | TTTCTCAGAGTGTTTGTGCTTAGT | MspJI,DdeI,MspJI,Hpy188I,DdeI,MspJI,MspJI,BspCNI,BseMII |
| 61 | dxz1 | ATCCGCAAGGGGATATTTGGACTT | AAGTCCAAATATCCCCTTGCGGAT | AciI,MspJI,MspJI,MspJI,MspJI |
| 62 | dxz1 | CCATAACTAAGCACAAACACTCTG | CAGAGTGTTTGTGCTTAGTTATGG | MspJI,FaiI,MspJI,DdeI,MspJI |
| 63 | dxz1 | AGCAGGTGGATATTTGGAGCTCTC | GAGAGCTCCAAATATCCACCTGCT | MspJI,SetI,SacI,Ecl136II,SduI,HgiJII*,HgiAI*,SetI,CviJI,AluI |
| 64 | dxz1 | TTCATTTGAAACGTCTATATCTTC | GAAGATATAGACGTTTCAAATGAA | AgsI,TaiI,SetI,MaeII,FaiI,MboII |
| 65 | dxz1 | CACACGTTTTGCAGAATCTGCAAG | CTTGCAGATTCTGCAAAACGTGTG | MspJI,AflIII,MspJI,TaiI,SetI,MaeII,CviRI*,MwoI,BstAPI,AlwNI,TfiI,HinfI,CviRI*,MspJI,MspJI |
| 66 | dxz1 | CTTTGTGATGTTTGTATTCAGCTC | GAGCTGAATACAAACATCACAAAG | SetI,CviJI,AluI,MspJI |
| 67 | dxz1 | CATTCACCTGACAGAGTTGAACTT | AAGTTCAACTCTGTCAGGTGAATG | SetI,MspJI,AgsI |
| 68 | dxz1 | AACCTGCCTTTGAGAGTTCATGTT | AACATGAACTCTCAAAGGCAGGTT | SetI,BspMI,MspJI,LpnPI,FatI,FaiI,NlaIII,CviAII,TspDTI,MspJI |
| 69 | dxz1 | TTCCCATAACTAAGCACAAACACT | AGTGTTTGTGCTTAGTTATGGGAA | MspJI,MspJI,FaiI,DdeI |
| 70 | dxz1 | GGAATAATTTCCCATAACTAAGCA | TGCTTAGTTATGGGAAATTATTCC | XmnI,TspEI*,FaiI,DdeI,MspJI |
| 71 | dxz1 | GATTTCTTTCGAAACGGGAATATT | AATATTCCCGTTTCGAAAGAAATC | AsuII,TaqI,MspJI,SspI |
| 72 | dxz1 | GCCAGTGGTAGTGAAGGAAAGAAC | GTTCTTTCCTTCACTACCACTGGC | MspJI,LpnPI,BtsIMutI,Hin4II* |
| 73 | dxz1 | GAAGCATTCTCAGAAACTTGTGAG | CTCACAAGTTTCTGAGAATGCTTC | DdeI,BspCNI,BsmI,BseMII,Hpy188I,MspJI,MspJI,MspJI |
| 74 | dxz1 | TTTGTAGCCTTCGTTTGAAACATC | GATGTTTCAAACGAAGGCTACAAA | CviJI,Hin4II*,AgsI |
| 75 | dxz1 | GCATTCTCAGAAACTTGTGAGTGA | TCACTCACAAGTTTCTGAGAATGC | MspJI,DdeI,BspCNI,BseMII,Hpy188I,MspJI,MspJI |
| 76 | dxz1 | GCGGACTTGGAGGACTGTGTTGGA | TCCAACACAGTCCTCCAAGTCCGC | MspJI,AciI,MspJI,MnlI,Tsp4CI*,MspJI,MspJI |
| 77 | dxz1 | GTTTGCATTCACCTGACAGAGTTG | CAACTCTGTCAGGTGAATGCAAAC | CviRI*,BsmI,SetI |
| 78 | dxz1 | CGTTGGAAAAGGGAATAATTTCCC | GGGAAATTATTCCCTTTTCCAACG | XmnI,TspEI* |
| 79 | dxz1 | ACTCCCAGGGTTGAACATTCCTTT | AAAGGAATGTTCAACCCTGGGAGT | PasI,MspJI,SecI*,MspJI,LpnPI,BssKI,SecI*,ScrFI,BciT130I,EcoRII,MspJI,AgsI |
| 80 | dxz1 | GACGGAGTTTCACTCACAGAGCTG | CAGCTCTGTGAGTGAAACTCCGTC | TspGWI,MspJI,SetI,CviJI,AluI |
| 81 | dxz1 | TTCTTTGGAAACGGGATCAACTTC | GAAGTTGATCCCGTTTCCAAAGAA | BinI*,DpnI,BstKTI,MboI |
| 82 | dxz1 | TAGAATCTGCAAATGGAGATTTGC | GCAAATCTCCATTTGCAGATTCTA | TfiI,HinfI,CviRI*,MspJI |
| 83 | dxz1 | CCAAATACACTTTTGGTAGAATCA | TGATTCTACCAAAAGTGTATTTGG | MspJI,MspJI,TfiI,HinfI |
| 84 | dxz1 | ATGAATGCATTCAAGTCACACAGT | ACTGTGTGACTTGAATGCATTCAT | TspDTI,BsmI,XmnI,EcoT22I,CviRI*,AgsI,BsmI,Tsp45I,MaeIII,Tsp4CI* |
| 85 | dxz1 | GGGCTTGGAGGATTGTGTTGGAAA | TTTCCAACACAATCCTCCAAGCCC | CviJI,MspJI,MspJI |
| 86 | dxz1 | TCTCAGAAACTTCTGAGTGATGAA | TTCATCACTCAGAAGTTTCTGAGA | MspJI,DdeI,BspCNI,BseMII,Hpy188I,MspJI,Hpy188I,DdeI,BspCNI,BseMII |
| 87 | dxz1 | GCATTCAACTCCCAGGGTTGAACA | TGTTCAACCCTGGGAGTTGAATGC | AgsI,PasI,SecI*,BssKI,SecI*,ScrFI,BciT130I,EcoRII,AgsI |
| 88 | dxz1 | CGGGAATAATTTCCCATACCTAAA | TTTAGGTATGGGAAATTATTCCCG | MspJI,XmnI,TspEI*,FaiI,SetI |
| 89 | dxz1 | CAACTCCCAGGGTTGAACATTCCT | AGGAATGTTCAACCCTGGGAGTTG | PasI,MspJI,SecI*,MspJI,BssKI,SecI*,ScrFI,BciT130I,EcoRII,AgsI |
| 90 | dxz1 | GGGATGTTTGCATTCACCTCTCAG | CTGAGAGGTGAATGCAAACATCCC | FokI,BseGI,CviRI*,BsmI,SetI,HphI,DdeI,MspJI |
| 91 | dxz1 | TCCCATAACTAAGCACAAACACTC | GAGTGTTTGTGCTTAGTTATGGGA | MspJI,MspJI,FaiI,DdeI |
| 92 | dxz1 | AAGCTAAACGGAAGCACTCTCAGA | TCTGAGAGTGCTTCCGTTTAGCTT | SetI,CviJI,AluI,MspJI,TspGWI,DdeI,Hpy188I,MspJI |
| 93 | dxz1 | AACACTCCTTTTCTGAGAATCTGCA | TGCAGATTCTCGAAAAGGAGTGTT | TaqI,Hpy178III*,TfiI,HinfI,CviRI*,MspJI |
| 94 | dxz1 | CCTGCCTTTGAGAGTTCAGGTTCA | TGAACCTGAACTCTCAAAGGCAGG | MspJI,LpnPI,SetI,MspJI,MspJI,LpnPI |
| 95 | dxz1 | CGTTTGGAGGGCTTGGAGGCCTGT | ACAGGCCTCCAAGCCCTCCAAACG | CviJI,StuI,CviJI,HaeIII,MnlI,MspJI |
| 96 | dxz1 | TTGGTAGAATCAGCAGGTGGATAT | ATATCCACCTGCTGATTCTACCAA | TfiI,HinfI,SetI,BspMI,AarI |
| 97 | dxz1 | AATAATTTCCCATACCTAAACACA | TGTGTTTAGGTATGGGAAATTATT | TspEI*,FaiI,SetI |
| 98 | dxz1 | CTGAGTGATGAATGCATTCAAGTC | GACTTGAATGCATTCATCACTCAG | MspJI,DdeI,TspDTI,BsmI,XmnI,EcoT22I,CviRI*,AgsI,BsmI,MspJI |
| 99 | dxz1 | ATGAAACACTCCTTTTCTGAGAATC | GATTCTCGAAAAGGAGTGTTTCAT | TspDTI,TaqI,Hpy178III*,MspJI,TfiI,MspJI,HinfI |
| 100 | dxz1 | TGATGATTGCATTCAACTCCCAGG | CCTGGGAGTTGAATGCAATCATCA | CviRI*,AgsI,BsmI,SecI*,ScrFI,BciT130I,MspJI,MspJI,LpnPI |
| 101 | dxz1 | CTGAAGCATTCTCAGAAACGGCTT | AAGCCGTTTCTGAGAATGCTTCAG | MspJI,Eco57I,DdeI,BsmI,Hpy188I,CviJI |
| 102 | dxz1 | AAGATTTCTTTCGAAACGGGAATA | TATTCCCGTTTCGAAAGAAATCTT | AsuII,TaqI,MspJI |
| 103 | dxz1 | AGCTAAACGGAAGCACTCTCAGAA | TTCTGAGAGTGCTTCCGTTTAGCT | CviJI,AluI,MspJI,TspGWI,DdeI,Hpy188I,MspJI |
| 104 | dxz1 | GATGTTTGCATTCACCTGACAGAG | CTCTGTCAGGTGAATGCAAACATC | CviRI*,BsmI,SetI,HphI,MspJI |
| 105 | dxz1 | CCTTCGTTTGAAACATCTATATCT | AGATATAGATGTTTCAAACGAAGG | Hin4II*,AgsI,FaiI |
| 106 | dxz1 | GATTGCATTCAACTCCCAGGGTTG | CAACCCTGGGAGTTGAATGCAATC | CviRI*,AgsI,BsmI,PasI,SecI*,BssKI,SecI*,ScrFI,BciT130I,EcoRII,MspJI,MspJI,LpnPI |
| 107 | dxz1 | CTCCCAGGGTTGAACATTCCTTTT | AAAAGGAATGTTCAACCCTGGGAG | PasI,MspJI,SecI*,MspJI,LpnPI,BssKI,SecI*,ScrFI,BciT130I,EcoRII,MspJI,AgsI |
| 108 | dxz1 | GATTTCTTTGGAAACGGGATCAAC | GTTGATCCCGTTTCCAAAGAAATC | MspJI,DpnI,BstKTI,MboI |
| 109 | dxz1 | CTTCTGATGGAGCAGTTTTTAAAC | GTTTAAAAACTGCTCCATCAGAAG | Hpy188I,MspJI,AhaIII*,MseI |
| 110 | dxz1 | CCAGTGGTAGTGAAGGAAAGAACT | AGTTCTTTCCTTCACTACCACTGG | MspJI,LpnPI,BtsIMutI,Hin4II* |
| 111 | dxz1 | GCAAATGGAGATTTGCACTGCTTT | AAAGCAGTGCAAATCTCCATTTGC | MspJI,CviRI*,TspRI,BtsIMutI,BtsI |
| 112 | dxz1 | GATTTCGTTGGAAAAGGGAATAAT | ATTATTCCCTTTTCCAACGAAATC | |
| 113 | dxz1 | TGAGAGTTCAGGTTCAAAACACTC | GAGTGTTTTGAACCTGAACTCTCA | SetI,AgsI |
| 114 | dxz1 | ATAATTTCCCATAACTAAGCACAA | TTGTGCTTAGTTATGGGAAATTAT | TspEI*,FaiI,DdeI,MspJI |
| 115 | dxz1 | AGGTTCAAAACACTCTTTCTGTAG | CTACAGAAAGAGTGTTTTGAACCT | AgsI,MspJI,SfeI*,MspJI |
| 116 | dxz1 | CTGCCTTTGAGAGTTCAGGTTCAA | TTGAACCTGAACTCTCAAAGGCAG | SetI,MspJI,MspJI,LpnPI,AgsI |
| 117 | dxz1 | CCTCTCAGAGTTGAACTTTCCCTT | AAGGGAAAGTTCAACTCTGAGAGG | MnlI,MspJI,DdeI,BspCNI,BseMII,Hpy188I,MspJI,AgsI |
| 118 | dxz1 | TCTGCAAATGGAGATTTGCACTGC | GCAGTGCAAATCTCCATTTGCAGA | CviRI*,MspJI,CviRI*,MspJI,BtsIMutI,BtsI |
| 119 | dxz1 | TCAGCAGGTGGATATTTGGAGCTC | GAGCTCCAAATATCCACCTGCTGA | MspJI,SetI,MspJI,SacI,MspJI,Ecl136II,SduI,HgiJII*,HgiAI*,SetI,CviJI,AluI |
| 120 | dxz1 | GGCCAGTGGTAGTGAAGGAAAGAA | TTCTTTCCTTCACTACCACTGGCC | CviJI,HaeIII,MspJI,LpnPI,TspRI,BtsIMutI,BsrI,Hin4II* |
| 121 | dxz1 | AACTTCATATAAAAGGCAAATGGA | TCCATTTGCCTTTTATATGAAGTT | MspJI,FaiI,FaiI |
| 122 | dxz1 | AAATACACTTTTGGTAGAATCAGC | GCTGATTCTACCAAAAGTGTATTT | TfiI,HinfI,MspJI |
| 123 | dxz1 | TTTCTTTGGAAACGGGATCAACTT | AAGTTGATCCCGTTTCCAAAGAAA | DpnI,BstKTI,MboI |
| 124 | dxz1 | AATTTCCCATAACTAAGCACAAAC | GTTTGTGCTTAGTTATGGGAAATT | MspJI,FaiI,DdeI |
| 125 | dxz1 | TCATGGAGTTGAACAGTCCTATTG | CAATAGGACTGTTCAACTCCATGA | MspJI,FatI,FaiI,NlaIII,CviAII,AgsI,Tsp4CI* |
| 126 | dxz1 | TGGAAAAGGGAATAATTTCCCATA | TATGGGAAATTATTCCCTTTTCCA | XmnI,TspEI*,FaiI |
| 127 | dxz1 | TTGGACTTCTTTGAAGGTTTCGTT | AACGAAACCTTCAAAGAAGTCCAA | AgsI,SetI,Hin4II*,MspJI |
| 128 | dxz1 | AAAAGGGAATAATTTCCCATAACT | AGTTATGGGAAATTATTCCCTTTT | XmnI,TspEI*,FaiI |
| 129 | dxz1 | ACGGAGTTTCACTCACAGAGCTGA | TCAGCTCTGTGAGTGAAACTCCGT | TspGWI,MspJI,SetI,CviJI,AluI |
| 130 | dxz1 | GATGAATGCATTTAACTTGCAGAG | CTCTGCAAGTTAAATGCATTCATC | TspDTI,BsmI,EcoT22I,CviRI*,MseI,CviRI*,MspJI |
| 131 | dxz1 | GCAGATTGGAATCACTCTTTTTAT | ATAAAAAGAGTGATTCCAATCTGC | MspJI,BsaBI,TfiI,HinfI |
| 132 | dxz1 | ATTTGAAACGTCTATATCTTCACA | TGTGAAGATATAGACGTTTCAAAT | AgsI,TaiI,SetI,MaeII,FaiI,MboII |
| 133 | dxz1 | GAAACTTCTGAGTGATGAATGCAT | ATGCATTCATCACTCAGAAGTTTC | Hpy188I,DdeI,BsmI,EcoT22I,CviRI* |
| 134 | dxz1 | CGGGAATAATTTCCCATAACTAAG | CTTAGTTATGGGAAATTATTCCCG | MspJI,XmnI,TspEI*,FaiI,DdeI,MspJI |
| 135 | dxz1 | AACTAAGCACAAACACTCTGAGAA | TTCTCAGAGTGTTTGTGCTTAGTT | MspJI,DdeI,Hpy188I,DdeI,MspJI,MspJI,BspCNI,BseMII |
| 136 | dxz1 | TCAACTCATGGAGTTGAACAGTCC | GGACTGTTCAACTCCATGAGTTGA | MspJI,MspJI,FatI,FaiI,NlaIII,CviAII,AgsI,Tsp4CI* |
| 137 | dxz1 | ACGTTTTGCAGAATCTGCAAGGGG | CCCCTTGCAGATTCTGCAAAACGT | MaeII,CviRI*,MwoI,BstAPI,AlwNI,TfiI,HinfI,CviRI*,MspJI |
| 137 | dyz3 | TGTTCAACCCACAAAGTTGAACAT | ATGTTCAACTTTGTGGGTTGAACA | AgsI,AgsI |
| 138 | dyz3 | GATTGACAAGTTTTGAAAGACTAT | ATAGTCTTTCAAAACTTGTCAATC | MspJI,AgsI |
| 139 | dyz3 | GAGCCTTTTGTGGCCTACGGTGGG | CCCACCGTAGGCCACAAAAGGCTC | CviJI,CviJI,HaeIII,BsiYI*,Tsp4CI*,MspJI,MspJI |
| 140 | dyz3 | GTTTTAAGACCTAAGGTGGGAAAG | CTTTCCCACCTTAGGTCTTAAAAC | MseI,SetI,SauI*,BsiYI*,DdeI,SetI,MspJI |
| 141 | dyz3 | CAGAAGCTGAGAAACTTCTTTGTG | CACAAAGAAGTTTCTCAGCTTCTG | MspJI,AlwNI,SetI,CviJI,AluI,MspJI,DdeI,MspJI,MspJI |
| 142 | dyz3 | GATGTGTGCATTCATCAAACAGAA | TTCTGTTTGATGAATGCACACATC | CviRI*,BsmI |
| 143 | dyz3 | CGGAAATATGTTTACATAAAAACT | AGTTTTTATGTAAACATATTTCCG | MspJI,FaiI,Hpy166II,FaiI |
| 144 | dyz3 | GGTAGAAAAGAACTATCTTCACC | GGTGAAGATAGTTCTTTTTCTACC | MboII,HphI |
| 145 | dyz3 | GCATTCTGAGAAACTACCTTTTGA | TCAAAAGGTAGTTTCTCAGAATGC | Hpy188I,MspJI,DdeI,SetI,MspJI |
| 146 | dyz3 | TGGATATGGAGAGAGCTTTGAGGC | GCCTCAAAGCTCTCTCCATATCCA | FaiI,SetI,CviJI,AluI,MspJI,MspJI,MnlI |
| 147 | dyz3 | AGAAATATCTTCACAAAAATACTA | TAGTATTTTTGTGAAGATATTTCT | |
| 148 | dyz3 | CTTTATTTGATACAGCAGTTTTGA | TCAAAACTGCTGTATCAAATAAAG | MspJI |
| 149 | dyz3 | AACTACCCAGAAGCATTTTGAGAA | TTCTCAAAATGCTTCTGGGTAGTT | MspJI,MspJI,MspJI,MspJI |
| 150 | dyz3 | AGTTTGGAAGCACTCTTTCGGTAG | CTACCGAAAGAGTGCTTCCAAACT | MspJI,MspJI |
| 151 | dyz3 | TGGTTACTTGGAGACCTTTGTGGA | TCCACAAAGGTCTCCAAGTAACCA | MaeIII,MspJI,SetI,BsmAI,Eco31I,MspJI,MspJI |
| 152 | dyz3 | GCATTGGGAGAAAATTCTTGTGAT | ATCACAAGAATTTTCTCCCAATGC | ApoI,TspEI*,MspJI,MspJI |
| 153 | dyz3 | AAAACAACAAGGAAGCATTCTGAG | CTCAGAATGCTTCCTTGTTGTTTT | Hpy188I,DdeI,BsmI,MspJI,MspJI,BspCNI,BseMII |
| 154 | dyz3 | GAGGCCTGTTGTGGAAAACTAAAT | ATTTAGTTTTCCACAACAGGCCTC | StuI,CviJI,HaeIII,MspJI,LpnPI |
| 155 | dyz3 | ACTATTGTGGAAAAGGAATTATCT | AGATAATTCCTTTTCCACAATAGT | TspEI* |
| 156 | dyz3 | TGATTGACTAGTTTTGAAAATCTC | GAGATTTTCAAAACTAGTCAATCA | SpeI,MaeI,AgsI |
| 157 | dyz3 | TATCTTCACAAAAATACTACACAG | CTGTGTAGTATTTTTGTGAAGATA | MspJI |
| 158 | dyz3 | TACCTTTGGTAGAGCAGTTTTGAA | TTCAAAACTGCTCTACCAAAGGTA | SetI,AgsI,MspJI |
| 159 | dyz3 | TATGCATTCAACTCACAGAGACGA | TCGTCTCTGTGAGTTGAATGCATA | FaiI,EcoT22I,CviRI*,AgsI,BsmI,BsmAI,Esp3I |
| 160 | dyz3 | AAACTACACAGACGCATTCCGAGA | TCTCGGAATGCGTCTGTGTAGTTT | MspJI,Hpy188I,BsmI,MspJI,MspJI |
| 161 | dyz3 | TTTTTTTGTGCAGTTTTGAAACAA | TTGTTTCAAAACTGCACAAAAAAA | CviRI*,AgsI,MspJI |
| 162 | dyz3 | AGCAATCTTTTCTAGAATATGTA | TACATATTCTAGAAAAAGATTGCT | XbaI,Hpy178III*,MaeI,FaiI,MspJI |
| 163 | dyz3 | GACATGTGTGCACTAATCTCACAG | CTGTGAGATTAGTGCACACATGTC | BspLU11I*,NspI,AflIII,MspJI,FatI,FaiI,NlaIII,CviAII,Hpy166II,SduI,BseSI,ApaLI,HgiAI*,CviRI* |
| 164 | dyz3 | TATTTGGAGCCTTTTGGGTCTTAT | ATAAGACCCAAAAGGCTCCAAATA | NlaIV,CviJI,MspJI |
| 165 | dyz3 | GTGGTTTGAGACCTATGGTAGAAA | TTTCTACCATAGGTCTCAAACCAC | SetI,BsmAI,Eco31I,FaiI |
| 166 | dyz3 | AAGCAATCTTTTTCTAGAATATGT | ACATATTCTAGAAAAAGATTGCTT | XbaI,Hpy178III*,MaeI,FaiI,MspJI |
| 167 | dyz3 | CAAGTTTTGAAAGACTATGTTTCT | AGAAACATAGTCTTTCAAAACTTG | MspJI,AgsI,FaiI,MspJI |
| 168 | dyz3 | AGTTTAATCTATCATTTGATTGAG | CTCAATCAAATGATAGATTAAACT | MseI,MspJI,MspJI |
| 169 | dyz3 | AGGGAAAAGAAATGTCTTCACATA | TATGTGAAGACATTTCTTTTCCCT | MboII,BbvII*,FaiI |
| 170 | dyz3 | TCCAGGAGGATATTTGGAGTGCTT | AAGCACTCCAAATATCCTCCTGGA | PfoI,MspJI,LpnPI,BssKI,ScrFI,BciT130I,EcoRII,MspJI |
| 171 | dyz3 | CCTCACAGATTTGAATGTCTCTTT | AAAGAGACATTCAAATCTGTGAGG | MnlI,MspJI,MspJI,MspJI,AgsI |
| 172 | dyz3 | ATTCTGAGAAACCATTTTGTGCTG | CAGCACAAAATGGTTTCTCAGAAT | Hpy188I,MspJI,DdeI,MspJI,MspJI |
| 173 | dyz3 | TAAGACCTAAGGTGGGAAAGGAAA | TTTCCTTTCCCACCTTAGGTCTTA | SetI,MspJI,SauI*,BsiYI*,MspJI,DdeI,SetI |
| 174 | dyz3 | CTACCCAGAAGCATTTTGAGAAAC | GTTTCTCAAAATGCTTCTGGGTAG | MspJI,MspJI,LpnPI,MspJI,MspJI,MspJI |
| 175 | dyz3 | CCTTTGTGGCCCATGGTAGAAAAG | CTTTTCTACCATGGGCCACAAAGG | CviJI,HaeIII,AsuI*,DsaI*,StyI,NcoI,SecI*,FatI,FaiI,NlaIII,CviAII |
| 176 | dyz3 | GACTATGGAGGGAAAAGAAATGTC | GACATTTCTTTTCCCTCCATAGTC | FaiI |
| 177 | dyz3 | CACAAAAATACTACACAGAAGCAT | ATGCTTCTGTGTAGTATTTTTGTG | MspJI,MspJI |
| 178 | dyz3 | TTGTGGACTATGGAGGGAAAAGAA | TTCTTTTCCCTCCATAGTCCACAA | Hpy166II,FaiI,MnlI |
| 179 | dyz3 | TGTTTGTGATATGTACCTTCAACT | AGTTGAAGGTACATATCACAAACA | FaiI,Csp6I,RsaI,SetI,AgsI |
| 180 | dyz3 | AAAATTACACAGAGAGATTCTGAG | CTCAGAATCTCTCTGTGTAATTTT | TspEI*,TfiI,HinfI,Hpy188I,DdeI,MspJI,MspJI,BspCNI,BseMII |
| 181 | dyz3 | AGATGGTGGAAAAGGAAATGTCTT | AAGACATTTCCTTTTCCACCATCT | |
| 182 | dyz3 | AGCTTCTTTTTGATGTATGCATTC | GAATGCATACATCAAAAAGAAGCT | CviJI,AluI,FaiI,EcoT22I,CviRI*,MspJI,BsmI |
| 183 | dyz3 | AGAGCCATTTGTGGCCTATGGTGG | CCACCATAGGCCACAAATGGCTCT | CviJI,CviJI,HaeIII,FaiI,MspJI |
| 184 | dyz3 | GAAGCACTTTGAGGCCTATTGTTG | CAACAATAGGCCTCAAAGTGCTTC | StuI,CviJI,HaeIII,MnlI |
| 185 | dyz3 | AGTGGTTTGAGACCTATGGTAGAA | TTCTACCATAGGTCTCAAACCACT | SetI,BsmAI,Eco31I,FaiI,MspJI |
| 186 | dyz3 | GTATTCATACCACAGAGTCGAAAC | GTTTCGACTCTGTGGTATGAATAC | MspJI,FaiI,HinfI,TaqI |
| 187 | dyz3 | CGTTAAACTTACCTTTGGTAGAGC | GCTCTACCAAAGGTAAGTTTAACG | MseI,SetI,MspJI |
| 188 | dyz3 | GTTCAACCCACAAAGTTGAACATA | TATGTTCAACTTTGTGGGTTGAAC | AgsI,MspJI,AgsI,FaiI |
| 189 | dyz3 | GTTTGCAAACACTCTTTTTGTGGT | ACCACAAAAAGAGTGTTTGCAAAC | CviRI*,MspJI,MspJI,MspJI |
| 190 | dyz3 | CGGTAGAATCTGCAAGTGGATATG | CATATCCACTTGCAGATTCTACCG | TfiI,HinfI,CviRI*,FaiI,MspJI |
| 191 | dyz3 | GGAGACCTTTGTGGAAGATGGTGG | CCACCATCTTCCACAAAGGTCTCC | SetI,BccI |
| 192 | dyz3 | AAGAAATATCTTCATTTAAAAACT | AGTTTTTAAATGAAGATATTTCTT | AhaIII*,MseI |
| 193 | dyz3 | TAAAAACAACAAGGAAGCATTCTG | CAGAATGCTTCCTTGTTGTTTTTA | BsmI,MspJI |
| 194 | dyz3 | ACCTTTGTGGAAGATGGTGGAAAA | TTTTCCACCATCTTCCACAAAGGT | MboII,BccI |
| 195 | dyz3 | GCACTTTTCTGCCTATTGTGTAAA | TTTACACAATAGGCAGAAAAGTGC | MspJI |
| 196 | dyz3 | ATCATTTGATTGAGCAGTTTTAAA | TTTAAAACTGCTCAATCAAATGAT | AhaIII*,MseI |
| 197 | dyz3 | ATATTTGCAGCGCTTTGAGGCCTG | CAGGCCTCAAAGCGCTGCAAATAT | CviRI*,BlsI,BisI,BbvI,TseI,HaeII,Eco47III,Hin6I,GlaI,HhaI,StuI,MspJI,CviJI,HaeIII,MnlI,MspJI |
| 198 | dyz3 | GTTTTAAAAAACTTTTTTTGTGGA | TCCACAAAAAAGTTTTTTAAAAC | AhaIII*,MseI,MspJI,MspJI |
| 199 | dyz3 | TTCTTGTGATATTTGTGTTCAACC | GGTTGAACACAAATATCACAAGAA | MspJI,AgsI |
| 200 | dyz3 | GGAGGGAAAAGAAATGTCTTCACA | TGTGAAGACATTTCTTTTCCCTCC | MboII,BbvII* |
| 201 | dyz3 | AATTATCTTCTCATAAAACCTACA | TGTAGGTTTTATGAAGAGATAATT | FaiI,SetI |
| 202 | dyz3 | TATCGTTTGAGAGAGCATTTCGAA | TTCGAAATGCTCTCTCAAACGATA | AsuII,TaqI,MspJI |
| 203 | dyz3 | ATTCATCAAACAGAATTGAACATT | AATGTTCAATTCTGTTTGATGAAT | MspJI,TspEI*,AgsI |
| 204 | dyz3 | AAGGAAGCATTCTGAGAAACCATT | AATGGTTTCTCAGAATGCTTCCTT | Hpy188I,DdeI,BsmI,BspCNI,BseMII |
| 205 | dyz3 | TTGAATGTCTCTTTTGATTGAGCA | TGCTCAATCAAAAGAGACATTCAA | AgsI,BsmAI,MspJI |
| 206 | dyz3 | AAATGTCTTCCCGTAAAAACTACA | TGTAGTTTTACGGGAAGACATTT | |
| 207 | dyz3 | GAAACTTTTTTGACATGTGTGCAC | GTGCACACATGTCAAAAAAGTTTC | BspLU11I*,NspI,AflIII,FatI,FaiI,NlaIII,CviAII,MspJI,Hpy166II,SduI,BseSI,ApaLI,HgiAI*,CviRI* |
| 208 | dyz3 | GAAACATCTTCATATAAAAACAAC | GTTGTTTTTATATGAAGATGTTTC | FaiI,FaiI |
| 209 | dyz3 | TGTTTGATAGAGCAGTTGTGAAAC | GTTTCACAACTGCTCTATCAAACA | MspJI |
| 210 | dyz3 | CTTCATATAAAAACAACAAGGAAG | CTTCCTTGTTGTTTTTATATGAAG | MspJI,FaiI,FaiI,MspJI |
| 211 | dyz3 | ATTTAGAACGCTTGGAGGCCTATG | CATAGGCCTCCAAGCGTTCTAAAT | StuI,CviJI,HaeIII,MnlI,FaiI,MspJI |
| 212 | dyz3 | TAAAAGGAAATATCTTTACGTAAG | CTTACGTAAAGATATTTCCTTTTA | SnaBI,BsaAI,TaiI,SetI,MaeII,MspJI,MspJI |
| 213 | dyz3 | GTAGAAAAGGAACTATCTTCACAG | CTGTGAAGATAGTTCCTTTTCTAC | MboII |
| 214 | dyz3 | GCAAGTGGTTACTTGGAGACCTTT | AAAGGTCTCCAAGTAACCACTTGC | MspJI,MaeIII,SetI,BsmAI,Eco31I |
| 215 | dyz3 | GCGTTTTAAGACCTAAGGTGGGAA | TTCCCACCTTAGGTCTTAAAACGC | MseI,SetI,SauI*,BsiYI*,DdeI,SetI,MspJI |
| 216 | dyz3 | AATAAACTTGTTTGTGATATGTAC | GTACATATCACAAACAAGTTTATT | MspJI,FaiI,MspJI,Csp6I,RsaI |
| 217 | dyz3 | ATTTGGAGTGGTTTGAGACCTATG | CATAGGTCTCAAACCACTCCAAAT | SetI,BsmAI,Eco31I,FaiI,MspJI |
| 218 | dyz3 | GTGGATATTTAGAACGCTTGGAGG | CCTCCAAGCGTTCTAAATATCCAC | MspJI,MspJI,MnlI |
| 219 | dyz3 | GAAGTATCTTCCCTTAAAAGCTAT | ATAGCTTTTAAGGGAAGATACTTC | MseI,SetI,CviJI,AluI |
| 220 | dyz3 | CTATACAGAAGCGTTCCCAGAAAC | GTTTCTGGGAACGCTTCGTATAG | FaiI,MspJI,FaiI |
| 221 | dyz3 | ATTCTGAGAAACTTTTTTGACATG | CATGTCAAAAAAGTTTCTCAGAAT | Hpy188I,MspJI,DdeI,MspJI,FaiI,CviAII,MspJI |
| 222 | dyz3 | TGGCCCATGGTAGAAAAGGAACTA | TAGTTCCTTTTCTACCATGGGCCA | CviJI,HaeIII,AsuI*,MspJI,DsaI*,StyI,NcoI,SecI*,MspJI,FatI,FaiI,NlaIII,CviAII |
| 223 | dyz3 | TACTTGGAGACCTTTGTGGAAGAT | ATCTTCCACAAAGGTCTCCAAGTA | MspJI,SetI,BsmAI,Eco31I,MspJI |
| 224 | dyz3 | GTGGGTATTTAGAGCCATTTGTGG | CCACAAATGGCTCTAAATACCCAC | CviJI,MspJI,MspJI |
| 225 | dyz3 | AAAACTATACAGAAGCGTTCCCAG | CTGGGAACGCTTCTGTATAGTTTT | FaiI,MspJI |
| 226 | dyz3 | CTCTTTTCCTAGAATCTGTAAGTT | AACTTACAGATTCTAGGAAAAGAG | MaeI,TfiI,HinfI,MspJI,MspJI |
| 227 | dyz3 | GAGGCCTATGGTGCAAAAACGAAT | ATTCGTTTTTGCACCATAGGCCTC | StuI,CviJI,HaeIII,MwoI,MspJI,FaiI,CviRI* |
| 228 | dyz3 | CGAAACTATCGTTTGAGAGAGCAT | ATGCTCTCTCAAACGATAGTTTCG | MspJI |
| 229 | dyz3 | GAAGTGGATATTTAGAACGCTTGG | CCAAGCGTTCTAAATATCCACTTC | MspJI,MspJI |
| 230 | dyz3 | GAATATTTGGAGCCTTTTGGGTCT | AGACCCAAAAGGCTCCAAATATTC | SspI,NlaIV,CviJI,MspJI,MspJI,MspJI |
| 231 | dyz3 | TTCCCGTAAAAACTACACAGATGC | GCATCTGTGTAGTTTTTACGGGAA | MspJI,MspJI,SfaNI |
| 232 | dyz3 | AGGCCTATGGTGCAAAAACGAATA | TATTCGTTTTTGCACCATAGGCCT | StuI,CviJI,HaeIII,MwoI,MspJI,FaiI,CviRI* |
| 233 | dyz3 | CTTTTCATTGTGCAGTTTCCAAGC | GCTTGGAAACTGCACAATGAAAAG | CviRI*,MspJI |
| 234 | dyz3 | CTACACAGAAGCATTCCAATAAAC | GTTTATTGGAATGCTTCTGTGTAG | MspJI,MspJI,BsmI |
| 235 | dyz3 | GAAATATCTTCCCCTAAAAAGTAC | GTACTTTTTAGGGGAAGATATTTC | Csp6I,RsaI |
| 236 | dyz3 | ATATGGTGGAAAAGGAAACATCCG | CGGATGTTTCCTTTTCCACCATAT | FaiI,FokI,BseGI,MspJI |
| 237 | dyz3 | TATGTACCTTCAACTGACAGATTT | AAATCTGTCAGTTGAAGGTACATA | FaiI,Csp6I,RsaI,SetI,Hin4II*,AgsI |
| 238 | dyz3 | GCCTATTGTTGAAAAGGAAACATC | GATGTTTCCTTTTCAACAATAGGC | MspJI,AgsI |
| 239 | dyz3 | TTCACATAAACACTACTCAGAAGC | GCTTCTGAGTAGTGTTTATGTGAA | MspJI,MspJI,FaiI,DdeI,Hpy188I,MspJI |
| 240 | dyz3 | TGATATTTGTGTTCAACCCACAAA | TTTGTGGGTTGAACACAAATATCA | AgsI |
| 241 | dyz3 | AAATATCTTCCCCTAAAAAGTACA | TGTACTTTTTAGGGGAAGATATTT | TatI,Csp6I,RsaI |
| 242 | dyz3 | AACTATACAGAAGCGTTCCCAGAA | TTCTGGGAACGCTTCTGTATAGTT | FaiI,MspJI |
| 243 | dyz3 | TTTGAACATTCCTTTTTATAGAAT | ATTCTATAAAAAGGAATGTTCAAA | AgsI,FaiI |
| 244 | dyz3 | TATACAGAAGCGTTCCCAGAAACT | AGTTTCTGGGAACGCTTCTGTATA | FaiI,MspJI,MspJI |
| 245 | dyz3 | GCAGCGCTTTGAGGCCTGCGGTGG | CCACCGCAGGCCTCAAAGCGCTGC | BlsI,BisI,BbvI,TseI,HaeII,Eco47III,Hin6I,GlaI,HhaI,StuI,CviJI,HaeIII,MnlI,Cac8I,MspJI,MspJI,A |
| 246 | dyz3 | CTTGTGATATTTGTGTTCAACCCA | TGGGTTGAACACAAATATCACAAG | MspJI,AgsI |
| 247 | dyz3 | TTCGGAGCACTTTGAGGCCTGTTG | CAACAGGCCTCAAAGTGCTCCGAA | Hpy188I,MspJI,SduI,HgiAI*,StuI,CviJI,HaeIII,MnlI,MspJI |
| 248 | dyz3 | AACGCTTGGAGGCCTATGGTGCAA | TTGCACCATAGGCCTCCAAGCGTT | MspJI,StuI,CviJI,HaeIII,MnlI,MwoI,FaiI,MspJI,CviRI* |
| 249 | dyz3 | ATTTGTGTTCAACCCACAAAGTTG | CAACTTTGTGGGTTGAACACAAAT | AgsI |
| 250 | dyz3 | GAATCTGCAAGTGGATATGGAGAG | CTCTCCATATCCACTTGCAGATTC | TfiI,HinfI,CviRI*,FaiI,MspJI |
| 251 | dyz3 | AGCGTTTTAAGACCTAAGGTGGGA | TCCCACCTTAGGTCTTAAAACGCT | MseI,SetI,SauI*,BsiYI*,DdeI,SetI,MspJI,MspJI |
| 252 | dyz3 | GGTCTTATTGTGGAAAAGGAAATA | TATTTCCTTTTCCACAATAAGACC | MspJI |
| 253 | dyz3 | GAGAGATTCTGAGAAACTTCTTTG | CAAAGAAGTTTCTCAGAATCTCTC | TfiI,HinfI,Hpy188I,DdeI,MspJI |
| 254 | dyz3 | CGTTTTAAGACCTAAGGTGGGAAA | TTTCCCACCTTAGGTCTTAAAACG | MseI,SetI,SauI*,BsiYI*,DdeI,SetI,MspJI |
| 255 | dyz3 | GATATTTGGAGCACTTTTCTGCCT | AGGCAGAAAAGTGCTCCAAATATC | SduI,HgiAI*,MspJI |
| 256 | dyz3 | ATGGTGCAAAAACGAATAACTTCA | TGAAGTTATTCGTTTTTGCACCAT | CviRI*,MspJI,XmnI |
| 257 | dyz3 | AGAGATTCTGAGAAACTTCTTTGT | ACAAAGAAGTTTCTCAGAATCTCT | TfiI,HinfI,Hpy188I,DdeI,MspJI |
| 258 | dyz3 | GGAAAATGTAAGTGGGTATTTAGA | TCTAAATACCCACTTACATTTTCC | MspJI |
| 259 | dyz3 | GGAAAACTAAATGTCTTCATATAA | TTATATGAAGACATTTAGTTTTCC | MspJI,MboII,FaiI,BbvII*,TspDTI,FaiI |
| 260 | dyz3 | TACACAGAAGCATTGGGAGAAAAT | ATTTTCTCCCAATGCTTCTGTGTA | MspJI,MspJI |
| 261 | dyz3 | AAGTGGTTACTTGGAGACCTTTGT | ACAAAGGTCTCCAAGTAACCACTT | MaeIII,SetI,BsmAI,Eco31I,MspJI |
| 262 | dyz3 | AGTTCAACCTATCTTTTCGTAGAG | CTCTACGAAAAGATAGGTTGAACT | AgsI,SetI,MspJI |
| 263 | dyz3 | TGTGGAAAACTAAATGTCTTCATA | TATGAAGACATTTAGTTTTCCACA | MboII,FaiI,BbvII*,TspDTI |
| 264 | dyz3 | AGGCCTGTTGTGGAAAACTAAATG | CATTTAGTTTTCCACAACAGGCCT | StuI,CviJI,HaeIII,MspJI,LpnPI |
| 265 | dyz3 | CGGAGCACTTTGAGGCCTGTTGTG | CACAACAGGCCTCAAAGTGCTCCG | MspJI,SduI,HgiAI*,StuI,CviJI,HaeIII,MnlI,MspJI,MspJI |
| 266 | dyz3 | CATATAAAAACAACAAGGAAGCAT | ATGCTTCCTTGTTGTTTTTATATG | MspJI,FaiI,FaiI |
| 267 | dyz3 | GTTTGAGACCTATGGTAGAAAAAG | CTTTTTCTACCATAGGTCTCAAAC | SetI,FaiI |
| 268 | dyz3 | GGTAGAATCTGCAAGTGGATATGG | CCATATCCACTTGCAGATTCTACC | TfiI,HinfI,CviRI*,FaiI,MspJI |
| 269 | dyz3 | ACTATACAGAAGCGTTCCCAGAAA | TTTCTGGGAACGCTTCTGTATAGT | FaiI,MspJI,MspJI |
| 270 | dyz3 | CAGAGTTTAATCTATCATTTGATT | AATCAAATGATAGATTAAACTCTG | MspJI,MseI,MspJI |
| 271 | dyz3 | TCACCGAAAAACCACACAGAAGCA | TGCTTCTGTGTGGTTTTTCGGTGA | MspJI,MspJI |
| 272 | dyz3 | AACTACACAGACACATTCTGTGAA | TTCACAGAATGTGTCTGTGTAGTT | MspJI,MspJI,MspJI |
| 273 | dyz3 | TAAAAAACTTTTTTGTGGAATCT | AGATTCCACAAAAAAAGTTTTTTA | MspJI,TfiI,HinfI |
| 274 | dyz3 | AGAAGCTGAGAAACTTCTTTGTGA | TCACAAAGAAGTTTCTCAGCTTCT | SetI,CviJI,AluI,MspJI,DdeI,MspJI,MspJI |
| 275 | dyz3 | ATTCACCTCACAGATTTGAATGTC | GACATTCAAATCTGTGAGGTGAAT | SetI,MnlI,MspJI,AgsI,MspJI |
| 276 | dyz3 | TTTGTGTTCAACCCACAAAGTTGA | TCAACTTTGTGGGTTGAACACAAA | AgsI |
| 277 | dyz3 | CTTTTGATTGACTAGTTTTGAAAA | TTTTCAAAACTAGTCAATCAAAAG | SpeI,MaeI,AgsI,MspJI |
| 278 | dyz3 | AGAAAAGGAACTATCTTCACAGAA | TTCTGTGAAGATAGTTCCTTTTCT | MboII |
| 279 | dyz3 | CTATCGTTTGAGAGAGCATTTCGA | TCGAAATGCTCTCTCAAACGATAG | TaqI,MspJI |
| 280 | dyz3 | TTTTAAGACCTAAGGTGGGAAAGG | CCTTTCCCACCTTAGGTCTTAAAA | MseI,SetI,SauI*,BsiYI*,DdeI,SetI,MspJI |
| 281 | dyz3 | AATCTATCATTTGATTGAGCAGTT | AACTGCTCAATCAAATGATAGATT | MspJI |
| 282 | dyz3 | GAGCACTTTGTGGACTATGGAGGG | CCCTCCATAGTCCACAAAGTGCTC | SduI,HgiAI*,DraIII,Hpy166II,FaiI,MspJI,MnlI |
| 283 | dyz3 | TGAATATTTGGAGCCTTTTGGGTC | GACCCAAAAGGCTCCAAATATTCA | SspI,NlaIV,CviJI,MspJI,MspJI,MspJI |
| 284 | dyz3 | AAGTGGATATTTAGAACGCTTGGA | TCCAAGCGTTCTAAATATCCACTT | MspJI,MspJI |
| 285 | dyz3 | GTTTGATAGAGCAGTTGTGAAACT | AGTTTCACAACTGCTCTATCAAAC | MspJI |
| 286 | dyz3 | GACCTAAGGTGGGAAAGGAAATAT | ATATTTCCTTTCCCACCTTAGGTC | SetI,MspJI,SauI*,BsiYI*,MspJI,DdeI,SetI |
| 287 | dyz3 | GAGTTCAACCTATCTTTTCGTAGA | TCTACGAAAAGATAGGTTGAACTC | AgsI,SetI,MspJI |
| 288 | dyz3 | TAAACTTGTTTGTGATATGTACCT | AGGTACATATCACAAACAAGTTTA | MspJI,FaiI,MspJI,Csp6I,RsaI |
| 289 | dyz3 | TGGAAAACTAAATGTCTTCATATA | TATATGAAGACATTTAGTTTTCCA | MboII,FaiI,BbvII*,TspDTI,FaiI |
| 290 | dyz3 | GAAGCATTCCAATAAACTTGTTTG | CAAACAAGTTTATTGGAATGCTTC | BsmI,MspJI,MspJI |
| 291 | dyz3 | GAAATGTCTTCCCGTAAAAACTAC | GTAGTTTTTACGGGAAGACATTTC | |
| 292 | dyz3 | AAACCACAGAGTTGAACCTATCTT | AAGATAGGTTCAACTCTGTGGTTT | MspJI,MspJI,AgsI,SetI |
| 293 | dyz3 | AAACTATACAGAAGCGTTCCCAGA | TCTGGGAACGCTTCTGTATAGTTT | FaiI,MspJI |
| 294 | dyz3 | GAAAATGTAAGTGGGTATTTAGAG | CTCTAAATACCCACTTACATTTTC | MspJI |
| 295 | dyz3 | AGTGGTTACTTGGAGACCTTTGTG | CACAAAGGTCTCCAAGTAACCACT | MaeIII,SetI,BsmAI,Eco31I,MspJI,MspJI |
| 296 | dyz3 | AGAATCTAGAAGTGGATATTTAGA | TCTAAATATCCACTTCTAGATTCT | TfiI,HinfI,XbaI,Hpy178III*,MspJI,MaeI,MspJI |
| 297 | dyz3 | AAATGTCTTCATATAAAAGCTACA | TGTAGCTTTTATATGAAGACATTT | FaiI,FaiI,SetI,CviJI,AluI |
| 298 | dyz3 | ATGTGTGTATTTGTCTCAGACTGG | CCAGTCTGAGACAAATACACACAT | BsmAI,DdeI,Hpy188I,MspJI,MspJI,LpnPI |
| 299 | dyz3 | TCATAAAACCTACACTGAAGGATT | AATCCTTCAGTGTAGGTTTTATGA | MspJI,FaiI,SetI,TspRI,BtsIMutI,Hin4II* |
| 300 | dyz3 | CTAGTTTTGAAAATCTCTTTTTGT | ACAAAAGAGATTTTCAAAACTAG | MspJI,MaeI,AgsI,MspJI |
| 301 | dyz3 | TTTAGAACGCTTGGAGGCCTATGG | CCATAGGCCTCCAAGCGTTCTAAA | StuI,CviJI,HaeIII,MnlI,FaiI,MspJI |
| 302 | dyz3 | ACGAATAACTTCACACAAAAAATA | TATTTTTTGTGTGAAGTTATTCGT | MspJI,XmnI |